\documentclass[fleqn,usenatbib]{mnras}

\usepackage[T1]{fontenc}
\usepackage{graphicx}
\usepackage{xcolor}
\usepackage{color}
\usepackage{hyperref}
\usepackage{notes2bib}
\usepackage{float}
\usepackage{array}
\usepackage{graphicx} 
\usepackage{multirow}
\usepackage{hhline,booktabs}
\usepackage{makecell}
\usepackage{amsmath,amssymb}

\usepackage{newtxtext,newtxmath}

\hypersetup{colorlinks=true,linkcolor=[rgb]{1.,0.2,0.2},citecolor=[rgb]{0.1,0.4,1.},filecolor=[rgb]{0.7,0.2,0.2},urlcolor=[rgb]{0.0,0.2,1.}}
\makeatletter
\@fleqnfalse
\@mathmargin\@centering
\makeatother

\def\INSPIRE{\mbox{{\tt INSPIRE}}}
\newcommand{\Reff}{$\mathrm{R}_{\mathrm{e}\,}$}
\newcommand{\Mstar}{$\mathrm{M}_{\star}\,$}
\newcommand{\Mfrac}{$f_{M^{\star}_{t\text{BB}=3}}$}

\newcommand{\afe}{[$\alpha$/Fe]}
\newcommand{\kms}{km s$^{-1}$}
\newcommand{\Msun}{M$_{\odot}\,$}
\usepackage[normalem]{ulem}
\newcommand{\ppxf}{\textsc{pPXF}}

\definecolor{darkgreen}{rgb}{0.09, 0.45, 0.27}
\definecolor{amber(sae/ece)}{rgb}{1.0, 0.49, 0.0}
\defcitealias{Ferre-Mateu+17}{F17}
\defcitealias{Tortora+18_UCMGs}{T18}
\defcitealias{Scognamiglio20}{S20}
\defcitealias{Spiniello20_Pilot}{\INSPIRE\, Pilot}
\defcitealias{Spiniello+21}{\INSPIRE\, DR1}
\defcitealias{DAgo23}{\INSPIRE\, DR2}

\title[INSPIRE V. Relics and their degree of relicness]{ \centering INSPIRE: INvestigating Stellar Population In RElics V.\\ A catalogue of ultra-compact massive galaxies outside the local Universe and their degree of relicness}
\author[C. Spiniello et al.]{\noindent
C.~Spiniello$^{1, 2}$\thanks{E-mail: chiara.spiniello@physics.ox.ac.uk},
G.~D'Ago$^{3,2}$, 
L.~Coccato$^{4}$, J.~Hartke$^{5,6}$, 
C.~Tortora$^{2}$, 
A.~Ferr\'e-Mateu$^{7,8}$, 
C.~Pulsoni$^{9}$,\and
M.~Cappellari$^{1}$,
M.~Maksymowicz-Maciata$^{1}$, 
M.~Arnaboldi$^{4}$,
D.~Bevacqua$^{10,11}$, A.~Gallazzi$^{12}$, 
L.~K.~Hunt$^{12}$, \and
F.~La Barbera$^{2}$, 
I.~Mart\'in-Navarro$^{7,8}$, 
N.R.~Napolitano$^{13}$, 
M.~Radovich$^{14}$, 
P.~Saracco$^{15}$,
D.~Scognamiglio$^{16}$, \and
M.~Spavone$^{2}$, 
S.~Zibetti$^{12}$\\ 
$^{1}$Sub-Dep. of Astrophysics, Dep. of Physics, University of Oxford, Denys Wilkinson Building, Keble Road, Oxford OX1 3RH, United Kingdom\\
$^{2}$INAF -  Osservatorio Astronomico di Capodimonte, Via Moiariello  16, 80131, Naples, Italy\\
$^{3}$Institute of Astronomy, University of Cambridge, Madingley Road, Cambridge CB3 0HA, United Kingdom\\
$^{4}$European Southern Observatory,  Karl-Schwarzschild-Stra\ss{}e 2, 85748, Garching, Germany\\
$^{5}$Finnish Centre for Astronomy with ESO (FINCA), FI-20014 University of Turku, Finland\\
$^{6}$Tuorla Observatory, Department of Physics and Astronomy, FI-20014 University of Turku, Finland\\
$^{7}$Instituto de Astrof\'isica de Canarias, V\'ia L\'actea s/n, E-38205 La Laguna, Tenerife, Spain\\
$^{8}$Departamento de Astrofisica, Universidad de La Laguna, E-38200, La Laguna, Tenerife, Spain\\
$^{9}$Max-Planck-Institut f\"{u}r  extraterrestrische Physik, Giessenbachstrasse, 85748 Garching, Germany\\
$^{10}$INAF - Osservatorio Astronomico di Brera, via Brera 28, 20121 Milano, Italy\\
$^{11}$DiSAT, Universitá degli Studi dell’Insubria, via Valleggio 11, I-22100 Como, Italy\\
$^{12}$INAF - Osservatorio Astrofisico di Arcetri, Largo Enrico Fermi5, I-50125 Firenze, Italy\\
$^{13}$School for Physics and Astronomy, Sun Yat-sen University, Guangzhou 519082, Zhuhai Campus, China\\
$^{14}$INAF - Osservatorio astronomico di Padova, Vicolo Osservatorio 5, I-35122 Padova, Italy\\
$^{15}$INAF - Osservatorio Astronomico di Brera, via Brera 28, 20121 Milano, Italy\\
$^{16}$Jet Propulsion Laboratory, California Institute of Technology, 4800,  Oak Grove Drive - Pasadena, CA 91109, USA}
\date{Accepted XXX. Received YYY; in original form ZZZ}

\pubyear{2023}

\begin{document}
\label{firstpage}
\pagerange{\pageref{firstpage}--\pageref{lastpage}}
\maketitle
\begin{abstract}
%250 words. No references.
This paper presents the third data release of the INvestigating Stellar Population In RElics (INSPIRE) project, comprising 52 ultra-compact massive galaxies (UCMGs) observed with the X-Shooter spectrograph.  
We measure integrated stellar velocity dispersion, [Mg/Fe] abundances, ages, and metallicities for all the \INSPIRE\ objects. We thus infer star formation histories and confirm the existence of a {\sl degree of relicness} (DoR), defined in terms of the fraction of stellar mass formed by $z=2$, the time at which a galaxy has assembled 75\% of its mass, and the final assembly time. 
Objects with a high DoR assembled their stellar mass at early epochs, while low-DoR objects show a non-negligible fraction of later-formed populations and hence a spread in ages and metallicities. 
A higher DoR correlates with larger [Mg/Fe], 
super-solar metallicity, and larger velocity dispersion values. The 52 UMCGs span a large range of DoR from 0.83 to 0.06, with 38 of them having formed more than 75\% of their mass by $z=2$. 
Of these, nine are extreme relics (DoR$>0.7$), since they formed the totality ($>99\%$) of their stellar mass by redshift $z=2$. The remaining 14 UCMGs cannot be considered relics, as they are characterised by more extended star formation histories. 
With \INSPIRE\, we built the first sizeable sample of relics outside the local Universe, up to $z\sim0.4$, increasing the number of confirmed relics by a factor of $>10$, and opening up an important window to explain the mass assembly of massive galaxies in the high-z Universe. 
\end{abstract}


%%%%%%%%%%%%%%%%%%%%%%%%%%%%%%%%%%%%%%%%%%%%%%%%%%
%%%%%%%%%%%%%%%%% BODY OF PAPER %%%%%%%%%%%%%%%%%%

\section{Introduction}
Massive relics \citep{Trujillo+09_superdense} are nearby old ultra-compact massive galaxies (UCMGs) that descend from those high-redshift massive passive compact galaxies \citep{Nanayakkara+2022, Carnall+23}, the so-called “red nuggets”  \citep{Damjanov+09}.  
According to the two-phase formation scenario \citep{Naab+09, Naab+14, Oser+10, Huertas-Company+16}, red nuggets are formed at high redshift from an intense and fast dissipative series of processes that generate, after star formation (SF) quenches, a massive, passive and very compact galaxy. 
%which completely missed the merger and accretion phase predicted by the two-phase formation scenario \citep{Naab+09, Naab+14, Oser+10, Huertas-Company+16}. 
%This scenario predicts a first intense and fast dissipative series of processes that generate, after star formation (SF) quenches, a massive, passive and very compact galaxy (the so-called “red nuggets”, \citealt{Trujillo+09_superdense}. 
Then, after this dissipative phase is concluded (at $z\sim$2, \citealt{Zolotov15}), a second phase starts, characterised by gas inflows, dry mergers and interactions. This accretion phase, which is much more extended in time than the first one, is responsible for the dramatic structural evolution and size growth of ETGs from $z\sim2$ to today \citep{Buitrago+18_compacts}.

As the timescale for the initial phase is extremely short, observing massive galaxies at such stages is incredibly challenging. However, relics, 
slipping passively through cosmic time without interacting with any other structure, reached the nearby Universe as compact as they were when they formed and populated only by very ancient stars. Hence, as {\sl relics of the ancient Universe}, 
they allow us to study the physical processes that shaped the mass assembly of galaxies in the high-$z$ Universe with the amount of details only achievable in the nearby Universe. 

%Three relics have been confirmed so far in the local Universe (NGC~1277, Mrk~1216 and PGC~032873, \citealt{Trujillo+14, Ferre-Mateu+17}). They have been spectroscopically confirmed and fully characterised, in terms of their morpohological, photometrical, kinematical and stellar population properties \citep{Ferre-Mateu+17}. 
Since the discovery and spectroscopic confirmation of three relics in the local Universe \citep{Trujillo+14, Yildirim+15, Yildirim17,Ferre-Mateu+17}, a lot of work has been devoted to the search and confirmation of red UCMGs (e.g.~\citealt{Damjanov+15_compacts, Damjanov+15_env_compacts, Charbonnier+17_compact_galaxies, Tortora+16_compacts_KiDS, Tortora+18_UCMGs, Scognamiglio20}). These are the perfect \textit{'bonafide'} relic candidates, having red colours, very small sizes, and large stellar masses\footnote{Different works use different thresholds for the stellar mass and the half-light radius. Generally, the low-mass limit is a few times $10^{10}M_{\odot}$, and the upper size limit is $\sim 1.5-2.5$ kpc.}.
The number density of relics in the local Universe is very small (i.e., $\sim6 \times 10^{-7} \text{Mpc}^{-3}$, \citealt{Ferre-Mateu+17}). 
However, systematic surveys of UCMGs, have shown that their number density rises by almost two orders of magnitude between $z=0$ and $z\sim0.5$ \citep{Trujillo+12_compacts, Poggianti+13_low_z, Damjanov+13_compacts, Damjanov+14_compacts, Damjanov+15_compacts, Sanders14, Charbonnier+17_compact_galaxies, Tortora+18_UCMGs, Scognamiglio20,Lisiecki23}. This should imply that relics are 
more common at intermediate redshifts too, and hence, reliable statistical samples can be built at these redshifts. The main goal of the  INvestigating Stellar Population In RElics (\INSPIRE), described in the next section, is, indeed, to build the first statistically large sample of \textit{spectroscopically confirmed relics outside the local Universe}.  

The paper is organised as follows. The state-of-the-art in the topic and the current status of the \INSPIRE\ project, observations and previous results are described in Section~\ref{sec:inspire}, the data analysis and stellar kinematics are presented in Section~\ref{sec:data}, while the stellar population analysis is presented in Section~\ref{sec:stel_pop}. The main results, including the relic confirmation and characterisation, are presented in Section~\ref{sec:results}. Finally, we present our conclusions and outline the future development of \INSPIRE\ in  Section~\ref{sec:conclusions}. 

Throughout the paper, we assume a standard $\Lambda$CDM cosmology with $H_0$=$69.6$ \kms Mpc$^{-1}$, $\Omega_{\mathrm{\Lambda}}$=$0.714$ and $\Omega_{\mathrm{M}}$=$ 0.286$ \citep{Bennett14}.

\section{THE INSPIRE PROJECT}
\label{sec:inspire}
\INSPIRE\ builds up on a long process dedicated to  a systematic census of UCMGs in the redshift range $0.1<z<0.4$. 
First, we carried out a photometric selection of ultra-compact galaxies in the Kilo Degree Survey (KiDS, \citealt{Kuijken11}) DR3 footprint \citep{deJong+17_KiDS_DR3}, leveraging on its exquisite image quality (angular scale of $0.21\arcsec$/pixel and a median r-band seeing of $\sim0.65\arcsec$, \citealt{deJong+15_KiDS_paperI}) and wide-sky coverage \citep{Tortora+16_compacts_KiDS, Tortora+18_UCMGs}. 
Then, with a multi-site and multi-telescope spectroscopic follow-up campaign, we
confirmed $117$ of them,  obtaining low signal-to-noise (SNR), medium resolution optical spectra from which we infer the redshifts (\citealt{Tortora+18_UCMGs, Scognamiglio20}, hereafter T18 and S20, respectively). 
However, to finally confirm them as relics of the ancient Universe, it is necessary to constrain their stellar population parameters, demonstrating that the great majority of their stellar mass has assembled at very high redshift through a single, very short star formation episode. This is the main goal of the \INSPIRE\ Project, which aims 
%With the effort of enlarging the number of confirmed relics and also extend the redshift boundaries, we started the INvestigating Stellar Population In Relics (\INSPIRE) Project that aims  at 
at building the first large catalogue of spectroscopically confirmed relics at $0.1<z<0.4$. 
This redshift range represents a positive step forward in bridging the gap between the %an important link between 
very small sample of relics in the local Universe with the large population of high-redshift red nuggets. 

The \INSPIRE\, data have been collected as part of an ESO Large Programme (LP, ID: 1104.B-0370, PI: C. Spiniello) that started in P104 (October 2019) and was completed in March 2023 to obtain high SNR spectra with the X-Shooter spectrograph (XSH, \citealt{Vernet11}) of 52 UCMGs at redshift $0.1<z<0.4$. 
These objects all have 
%have been selected from a dedicated KiDS project (\citealt{Tortora+18_UCMGs,Scognamiglio20}, hereafter T18 and S20, respectively) and they have old stars as their 
$g-i$ broad band colour compatible with that of a stellar population with integrated ages $\ge8$ Gyr (considering a solar, super-solar and a sub-solar metallicity, see Fig.~1 in \citealt{Spiniello+21}). 
They also all have very small sizes (with effective radii \Reff$<2$ kpc), large stellar masses (\Mstar$>6\times10^{10}$\Msun), and are clear outliers in the stellar mass-size plane (see Fig.~2 in \citealt{Spiniello20_Pilot}).
Thanks to high SNR, wide wavelength spectra (covering from the UVB to the NIR), we are able to infer the stellar kinematics and population properties (age, metallicity, elemental abundance and low-mass end of the IMF slope) of these UCMGs, hence confirming or refuting their relic nature. 

\subsection{Previous Results}
In this section, we summarise the main results previously obtained by \INSPIRE, which have been presented in previous papers of the series. 
In \citet{Spiniello20_Pilot}, the Pilot Program of the Survey, we have presented 
several quantitative tests to validate our observational setup and strategy and our methods to infer kinematics and stellar population parameters. We have also obtained preliminary results on three systems, completely observed during the first semester of observations. 
With the first data release (\citealt{Spiniello+21}, hereafter INSPIRE DR1), we have carried out a precise stellar population analysis for 19 UCMGs. 
In particular, we have estimated the stellar population ages, metallicities and [Mg/Fe] abundances from their UVB and VIS spectra. Ten of them have been confirmed as relics (implying a preliminary successful rate for the survey of $\sim50$\%) since they had formed more than 75\% of their stellar masses early on in cosmic time through a star formation burst. 

%, confirming 50\% of them (10) as relics because they had formed 75\% or more of their stellar mass during the first phase of the formation scenario. 
The third paper of the series \citep{Martin-Navarro+23} focuses instead on the stellar Initial Mass Function (IMF), hinting at a possible dwarf-rich IMF in relics with respect to non-relics, although larger number statistics will be necessary to confirm this result. 

Finally, in DR2 \citet[hereafter INSPIRE DR2]{DAgo23}, the fourth paper of the series, we have performed a detailed stellar kinematic study on 40 objects, focusing on all sources of systematic uncertainties that affect the determination of the integrated stellar velocity dispersion. This quantity might be useful, in the future, as a selection criterion to identify reliable relic candidates. In fact, from DR1, we have shown that, at equal stellar mass, relics (and especially extreme relics) have larger velocity dispersion values than normal-sized ETGs and ultra-compact non-relics. However, this result needs to be confirmed with a larger statistical sample of galaxies. 

In \citetalias{Spiniello+21}, we adopted an operational definition of ``relic" as a way to identify objects that had formed the great majority of their stars at very high-z. 
Now that the \INSPIRE\ sample is completed, we move away from this empirical definition and  define instead a quantitative number, the ``degree of relicness" (DoR), for all the 52 UCMGs based on their star formation histories (SFH). As we will discuss in Section~\ref{sec:results}, the DoR not only provides a useful parameter to quantify how ``extreme'' the SFH is, enabling an operative definition of relics, but it also serves as a means to quantify the relative contribution of very old and lately formed (or accreted) stars. 
Finally, a larger number statistic, which is now achieved, will also allow us to assess whether the DoR correlates with morphological (sizes), photometrical (colours) and structural (boxyness/diskyness) properties and/or with the environment in which UCMGs live, and hence shed light on their formation mechanisms. 

%Within the pilot program and the first two data releases \citep{Spiniello20_Pilot, Spiniello+21, DAgo23}, we have analysed 40 out of 52 objects. In particular, we have presented several quantitative tests for the validity of our methods and strategy in the Pilot, have carried out a precise stellar population analysis for 19 of them in DR1, 

%A higher DoR indicates an earlier formation epoch with almost no contribution from stars brought in through accretion or formed in later SF episodes. 
%A lower DoR instead means that, although a fraction of stars are old and were formed during the first phase of the mass assembly, there is a non-negligible percentage of later-formed populations with different ages and metallicities. Hence, the DoR not only provides a useful parameter to quantify how ``extreme''  the SFH is, enabling an operative definition of relics, but it also serves as a means to quantify the relative contribution of pristine and lately formed (or accreted) stars. 

\subsection{Data Release 3: the complete sample}
\label{sec:inspiredr3}
%Here in this paper, we derive SFHs and obtain relic confirmation for 33 new systems: the 21 for which the stellar velocity dispersion has been constrained in DR2 and 12 new ones still unpublished, for which we also measure the stellar velocity dispersions. We also recompute the velocity dispersion and stellar population parameters of the 19 objects from DR1, in order to use the same exact configuration for all the objects. Hence, 
This paper presents the complete INSPIRE sample, comprising 52 UCMGs at $0.1<z<0.4$, all with \Reff$<2$ kpc and \Mstar$>6\times10^{10}$\Msun, of which 12 are new and still unpublished. 
For each UCMG, we have optical and NIR photometry from KiDS and VIKING \citep{Edge+14_VIKING-DR1}, 
structural parameters derived in \citet{Roy+18}, stellar masses 
inferred from SED fitting in the $ugri$ bands (\citetalias{Tortora+18_UCMGs, Scognamiglio20}). 
In this paper, we add to these morpho-photometric characteristics precise measurements of the stellar velocity dispersion, stellar age, metallicity, and [Mg/Fe] inferred from an integrated spectrum encapsulating 50\% of the light (R50, see Sec.~\ref{sec:data} for more info). 
All these quantities are released as part of a high-value catalogue to the ESO Science Portal.

\begin{table*}
\caption{Morpho-photometric characteristics of the INSPIRE final sample. Magnitudes ($gri$ in the AB system),  median effective radii (\Reff) in arcseconds and kpc, median S\'ersic indices ($n$), median axis ratios ($q$) and stellar masses have all been inferred from KiDS images.}
\label{tab:catalogue}
\begin{tabular}{lrrccrrrrrrrc}
\hline
\hline
  \multicolumn{1}{c}{ID} &
  \multicolumn{1}{c}{RA} &
  \multicolumn{1}{c}{DEC} &
  \multicolumn{1}{c}{DR} &
  \multicolumn{1}{c}{SAMPLE} &
  \multicolumn{1}{c}{mag$_{g}$} &
  \multicolumn{1}{c}{mag$_{r}$} &
  \multicolumn{1}{c}{mag$_{i}$} &
  \multicolumn{1}{c}{$\langle\mathrm{R}_{\mathrm{e}}\rangle\,$} &
  {$\langle\mathrm{R}_{\mathrm{e}}\rangle\,$} &
  \multicolumn{1}{c}{$\langle\mathrm{n}\rangle\,$} &    
  \multicolumn{1}{c}{$\langle\mathrm{q}\rangle\,$} &  
  \multicolumn{1}{c}{M$_{\star}$} \\

  \multicolumn{1}{c}{KiDS} &
  \multicolumn{1}{c}{(deg)} &
  \multicolumn{1}{c}{(deg)} &
  \multicolumn{1}{c}{} &
  \multicolumn{1}{c}{} &
  \multicolumn{1}{c}{(AB)} &
  \multicolumn{1}{c}{(AB)} &
  \multicolumn{1}{c}{(AB)} &
  \multicolumn{1}{c}{($\arcsec$)} &
\multicolumn{1}{c}{(kpc)} &
  \multicolumn{1}{c}{} &    
  \multicolumn{1}{c}{ } &  
  \multicolumn{1}{c}{($10^{11}$M$_{\odot}$)} \\
\hline
  J0211-3155 &  32.8962202  & -31.9279437  & 1 & T18       & 21.28 & 19.78 & 19.28 & 0.24 & 1.07  &  8.10 & 0.48 & 0.88 \\
  J0224-3143 &  36.0902655  & -31.7244923  & 1 & T18       & 20.91 & 19.25 & 18.62 & 0.29 & 1.55  &  6.06 & 0.39 & 2.71 \\
  J0226-3158 &  36.5109217  & -31.9810149  & 1 & T18       & 20.63 & 19.25 & 18.76 & 0.35 & 1.32  &  3.65 & 0.60 & 0.69 \\
  J0240-3141 &  40.0080971  & -31.6950406  & 1 & T18       & 20.58 & 19.05 & 18.59 & 0.19 & 0.81  &  8.10 & 0.27 & 0.98 \\
  J0314-3215 &  48.5942558  & -32.2632678  & 1 & T18       & 21.00 & 19.57 & 19.07 & 0.15 & 0.66  &  5.54 & 0.39 & 1.00 \\
  J0316-2953 &  49.1896388  & -29.8835868  & 1 & T18       & 21.19 & 19.66 & 19.13 & 0.20 & 1.02  &  3.52 & 0.31 & 0.87 \\
  J0317-2957 &  49.4141028  & -29.9561748  & 1 & T18       & 20.51 & 19.10 & 18.63 & 0.26 & 1.05  &  5.01 & 0.21 & 0.87 \\
  J0321-3213 &  50.2954390  & -32.2221290  & 1 & T18       & 20.67 & 19.23 & 18.74 & 0.31 & 1.37  &  4.93 & 0.39 & 1.23 \\
  J0326-3303 &  51.5140585  & -33.0540443  & 1 & T18       & 20.94 & 19.48 & 18.99 & 0.32 & 1.43  &  3.66 & 0.35 & 0.93 \\
  J0838+0052 & 129.5304520  &   0.8823841  & 1 & S20       & 20.65 & 19.29 & 18.75 & 0.31 & 1.28  &  4.02 & 0.41 & 0.87 \\
  J0842+0059 & 130.6665506  &   0.9899186  & 1 & S20       & 21.12 & 19.60 & 19.06 & 0.23 & 1.01  &  3.27 & 0.29 & 0.91 \\
  J0844+0148 & 131.0553886  &   1.8132204  & 2 & S20       & 21.25 & 19.78 & 19.26 & 0.26 & 1.14  &  6.56 & 0.49 & 0.71 \\
  J0847+0112 & 131.9112386  &   1.2057129  & 1 & SDSS-GAMA & 19.67 & 18.41 & 17.98 & 0.46 & 1.37  &  3.33 & 0.27 & 0.99 \\
  J0857-0108 & 134.2512185  &  -1.1457077  & 1 & S20       & 20.72 & 19.21 & 18.70 & 0.34 & 1.40  &  2.94 & 0.33 & 1.00 \\
  J0904-0018 & 136.0518949  &  -0.3054848  & 2 & S20       & 20.59 & 19.11 & 18.64 & 0.26 & 1.16  &  4.82 & 0.32 & 1.30 \\
  J0909+0147 & 137.3989150  &   1.7880025  & 2 & SDSS-GAMA & 20.06 & 18.68 & 18.16 & 0.30 & 1.05  &  9.97 & 0.77 & 1.05 \\
  J0917-0123 & 139.2701850  &  -1.3887918  & 2 & S20       & 20.86 & 19.21 & 18.66 & 0.27 & 1.37  &  3.05 & 0.41 & 2.19 \\
  J0918+0122 & 139.6446428  &   1.3794780  & 1 & T18       & 20.67 & 19.13 & 18.57 & 0.33 & 1.71  &  6.06 & 0.51 & 2.26 \\
  J0920+0126 & 140.1291393  &   1.4431610  & 2 & S20       & 20.97 & 19.52 & 19.05 & 0.33 & 1.51  &  6.92 & 0.68 & 0.98 \\
  J0920+0212 & 140.2320835  &   2.2126831  & 1 & SDSS-GAMA & 20.35 & 18.87 & 18.43 & 0.34 & 1.48  &  1.99 & 0.32 & 1.03 \\
  J1026+0033 & 156.7231818  &   0.5580980  & 2 & SDSS-GAMA & 18.45 & 17.39 & 16.97 & 0.34 & 1.02  &  3.18 & 0.29 & 1.48 \\
  J1040+0056 & 160.2152308  &   0.9407580  & 2 & S20       & 20.95 & 19.52 & 18.49 & 0.31 & 1.29  &  4.57 & 0.36 & 0.93 \\
  J1114+0039 & 168.6994335  &   0.6510299  & 2 & S20       & 20.45 & 19.00 & 18.55 & 0.34 & 1.52  &  4.93 & 0.25 & 1.62 \\
  J1128-0153 & 172.0885023  &  -1.8890642  & 2 & T18       & 19.87 & 18.56 & 18.07 & 0.35 & 1.27  &  6.69 & 0.31 & 1.30 \\
  J1142+0012 & 175.7023296  &   0.2043419  & 2 & SDSS-GAMA & 17.80 & 17.02 & 16.57 & 0.71 & 1.40  &  3.60 & 0.23 & 0.84 \\
  J1154-0016 & 178.6922829  &  -0.2779248  & 2 & T18       & 20.90 & 19.52 & 18.73 & 0.22 & 1.06  &  4.36 & 0.19 & 0.64 \\
  J1156-0023 & 179.2186145  &  -0.3946596  & 2 & SDSS-GAMA & 20.06 & 18.83 & 18.08 & 0.26 & 1.04  &  6.53 & 0.38 & 1.39 \\
  J1202+0251 & 180.5132277  &   2.8515451  & 2 & S20       & 20.97 & 19.43 & 18.96 & 0.31 & 1.49  &  6.47 & 0.89 & 0.68 \\
  J1218+0232 & 184.7355807  &   2.5449139  & 2 & S20       & 20.78 & 19.23 & 18.71 & 0.31 & 1.40  &  2.75 & 0.26 & 0.93 \\
  J1228-0153 & 187.0640987  &  -1.8989049  & 2 & S20       & 20.27 & 18.85 & 18.37 & 0.36 & 1.61  &  2.87 & 0.54 & 1.15 \\  J1402+0117 & 210.7400749  &   1.2917747  & 3 & S20       & 21.34 & 19.96 & 19.44 & 0.17 & 0.68  &  6.43 & 0.46 & 0.66 \\
  J1411+0233 & 212.8336012  &   2.5618381  & 2 & S20       & 20.49 & 18.86 & 18.41 & 0.21 & 1.07  &  2.83 & 0.30 & 1.55 \\
  J1412-0020 & 213.0038281  &  -0.3440699  & 3 & SDSS-GAMA & 20.74 & 19.19 & 18.67 & 0.33 & 1.42  &  6.13 & 0.39 & 1.20 \\
  J1414+0004 & 213.5646898  &   0.0809744  & 3 & SDSS-GAMA & 20.41 & 18.99 & 18.50 & 0.31 & 1.42  &  4.26 & 0.42 & 1.18 \\
  J1417+0106 & 214.3685124  &   1.1073909  & 3 & SDSS-GAMA & 18.97 & 17.90 & 17.51 & 0.49 & 1.48  &  3.92 & 0.33 & 0.91 \\
  J1420-0035 & 215.1715599  &  -0.5864629  & 3 & SDSS-GAMA & 20.30 & 18.95 & 18.45 & 0.34 & 1.35  &  5.67 & 0.62 & 0.99 \\
  J1436+0007 & 219.0481314  &   0.1217459  & 2 & SDSS-GAMA & 19.55 & 18.27 & 17.85 & 0.39 & 1.40  &  2.65 & 0.19 & 1.15 \\
  J1438-0127 & 219.5218882  &  -1.4582727  & 3 & SDSS-GAMA & 20.65 & 19.29 & 18.74 & 0.28 & 1.20  &  4.11 & 0.38 & 0.88 \\
  J1447-0149 & 221.9657402  &  -1.8242806  & 3 & SDSS-GAMA & 19.86 & 18.61 & 18.16 & 0.44 & 1.51  &  3.06 & 0.45 & 0.86 \\
  J1449-0138 & 222.3504660  &  -1.6459975  & 3 & SDSS-GAMA & 20.86 & 19.40 & 18.92 & 0.35 & 1.44  &  5.81 & 0.33 & 1.03 \\
  J1456+0020 & 224.2361596  &   0.3353906  & 3 & S20       & 20.88 & 19.46 & 18.97 & 0.12 & 0.50  &  5.53 & 0.20 & 0.71 \\
  J1457-0140 & 224.3397592  &  -1.6691725  & 3 & S20       & 20.99 & 19.43 & 18.93 & 0.34 & 1.66  &  4.60 & 0.53 & 1.51 \\
  J1527-0012 & 231.7772381  &  -0.2065670  & 3 & S20       & 21.37 & 19.67 & 19.08 & 0.23 & 1.26  &  5.77 & 0.23 & 1.74 \\
  J1527-0023 & 231.7522351  &  -0.3997483  & 3 & S20       & 21.21 & 19.64 & 19.14 & 0.22 & 1.12  &  9.16 & 0.75 & 1.15 \\
  J2202-3101 & 330.5472803  &  -31.018381  & 2 & T18       & 20.93 & 19.43 & 18.93 & 0.31 & 1.45  &  4.24 & 0.39 & 1.10 \\
  J2204-3112 & 331.2228147  &  -31.200261  & 2 & SDSS-GAMA & 20.84 & 19.32 & 18.86 & 0.35 & 1.39  &  6.36 & 0.31 & 0.90 \\
  J2257-3306 & 344.3966471  &  -33.114445  & 2 & T18       & 20.80 & 19.42 & 18.95 & 0.29 & 1.18  &  4.31 & 0.41 & 0.93 \\
  J2305-3436 & 346.3356634  &  -34.603091  & 1 & T18       & 21.26 & 19.69 & 19.12 & 0.31 & 1.29  &  3.89 & 0.40 & 0.86 \\
  J2312-3438 & 348.2389042  &  -34.648591  & 1 & T18       & 20.90 & 19.32 & 18.79 & 0.24 & 1.25  &  2.25 & 0.43 & 1.34 \\
  J2327-3312 & 351.9910156  &  -33.200760  & 1 & T18       & 20.99 & 19.35 & 18.80 & 0.28 & 1.51  &  5.94 & 0.67 & 1.57 \\
  J2356-3332 & 359.1261248  &  -33.533475  & 2 & T18       & 21.31 & 19.81 & 19.27 & 0.22 & 1.06  &  4.28 & 0.34 & 0.98 \\
  J2359-3320 & 359.9851685  &  -33.333583  & 1 & T18       & 21.11 & 19.59 & 19.05 & 0.24 & 1.04  &  4.49 & 0.39 & 1.07 \\
\hline
\hline\end{tabular}
%\begin{flushright}
%\footnotesize{$^*$ the objects have been taken %from SDSS or GAMA.}
%\end{flushright}
\end{table*}

Table~\ref{tab:catalogue} gives a summary of the photometric and structural characteristics of the complete \INSPIRE\ sample. These are taken from \citetalias{Tortora+18_UCMGs} and \citetalias{Scognamiglio20} and based on the analysis of optical images from the KiDS Survey, as detailed below. 
Specifically, together with ID and coordinates, we list the INSPIRE Data Release (DR) where the UCMG has been first presented and the sample from which it is drawn. Then, we provide $g$-, $r$- and $i$-band magnitudes, corrected by extinction, the circularised effective radii (\Reff), S\'ersic index ($n$), and axis ratio ($q$). These last three quantities have been calculated as the median of the quantities inferred from $g$, $r$ and $i$ images. 
Measures in each single band are reported in  \citetalias{Tortora+18_UCMGs} and \citetalias{Scognamiglio20} for all 52 systems. They are obtained by fitting a point-spread function (PSF) convolved S\'ersic profile to the images using the code \textsc{2dphot} \citep{LaBarbera_08_2DPHOT}\footnote{The source code of the package is available on request to the authors.}.  
%The effective radii are also translated into kiloparsecs using a Python version of the Ned Wright's Cosmology Calculator\footnote{\url{http://www.astro.ucla.edu/~wright/CosmoCalc.html}}. 
Uncertainties on the sizes, given the range covered by the \INSPIRE\, objects, are of the order of 20\% for objects with effective radii larger than 0.2 but rise up to more than 50\% for even smaller objects, as visible from Figure~B2 in \citetalias{Tortora+18_UCMGs}. We refer the reader to the same paper, where the quality of these structural parameters is assessed through simulated images.  
Finally, the last column of the table lists stellar masses, also taken from \citetalias{Tortora+18_UCMGs} and \citetalias{Scognamiglio20} and inferred from SED fitting in the $ugri$ bands.

\begin{figure*}
    \centering
    \includegraphics[width=18cm]{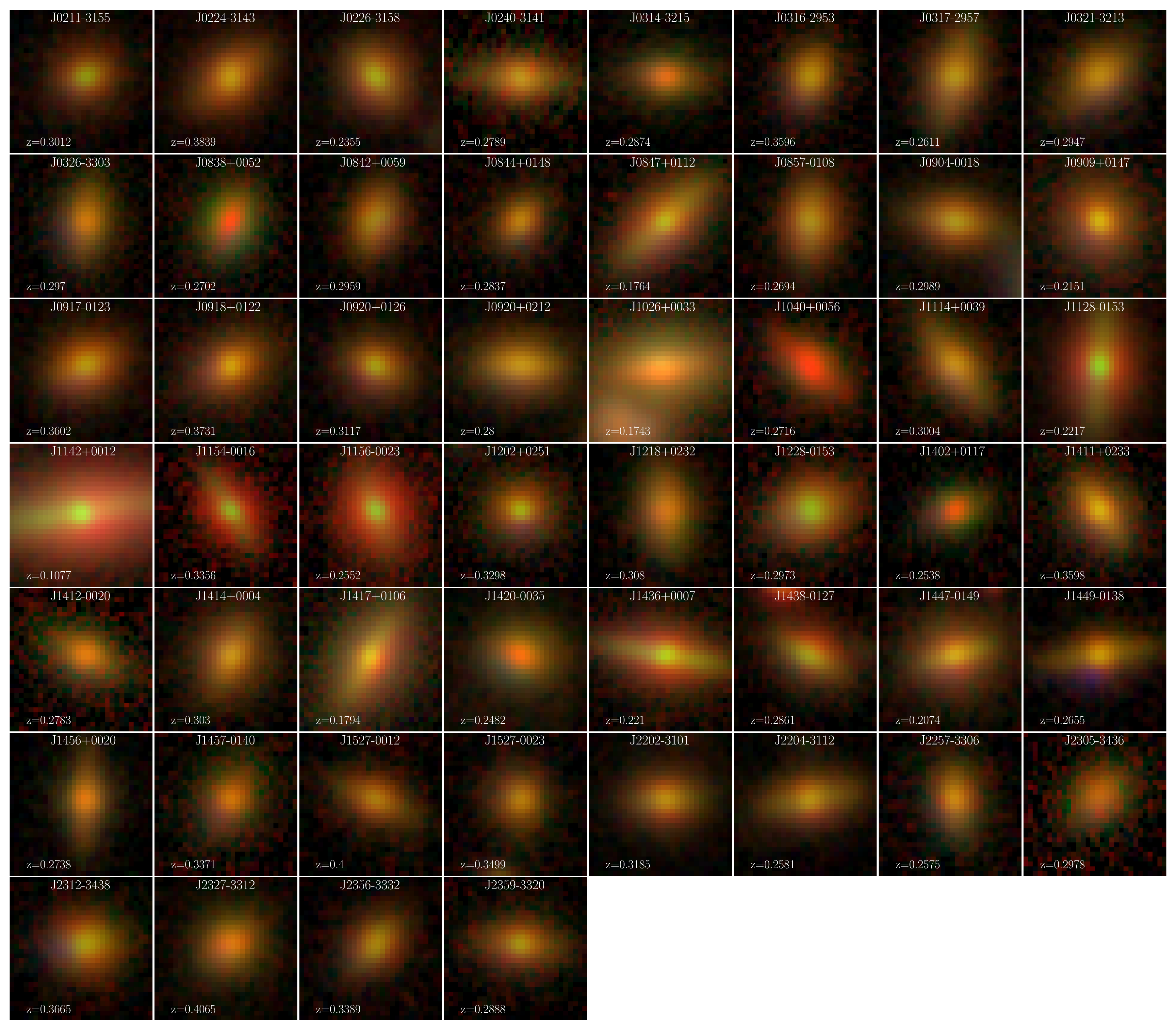}
    \caption{Colour-composite $gri$ KiDS images of the 52 \INSPIRE\ galaxies. Each panel has dimension of $6\arcsec\times6\arcsec$ and shows the ID of the corresponding galaxy on the top and its redshift on the bottom-left corner. For more details on astrometry and centering, see the text (Sec.~\ref{sec:inspiredr3}).}
    %Cutouts of the 52 \INSPIRE\ galaxies. Each panel shows a $5\arcsec\times5\arcsec$ $r-$band image from the KiDS survey. The major (Ma) and minor (mi) axis radii of the blue ellipses are calculated with the  circularised effective radii and the axis ratios, listed in Table~\ref{tab:catalogue}.} %The galaxy having a close by companion is J1026+0033, we note however that the second object is not present in the XSH slit. }
    \label{fig:cutouts}
\end{figure*}

Figure~\ref{fig:cutouts} shows $6\arcsec\times6\arcsec$ cutouts obtained combining the $g$, $r$ and $i$ bands from KiDS images. 
We note that the astrometry is not perfect in few cases. We have thus realigned the galaxy centers by fitting them with a Gaussian. 
%The blue ellipse in each panel encloses half of the total light of the galaxies. The major (Ma) and minor (mi) axis radius of each ellipse are calculated from the circularised effective radii and the axis ratios\footnote{R$_{\text{Ma}}=$\Reff$/\sqrt{q}$,  and R$_{\rm mi} =$R$_{\rm Ma}\times q$}. 
The galaxies appear all very small and compact on the sky, although in many cases hints of a flattened edge on structure can be seen\footnote{In the particular case of J1026+0033, a second galaxy can be seen in the field of view. However, we stress that this object does not fit into the X-Shooter slit and hence does not influence the results presented in this paper.}. Unfortunately, since the point-spread funtion of KiDS images is comparable to the sizes of the \INSPIRE\ galaxies, it is impossible to obtain a detailed morphological classification. Obtaining photometric data with better spatial sampling and higher resolution, either via AO supported imaging or from space, is the only way to infer more precise structural parameters for these ultra-compact galaxies.

%A high-level catalogue with all structural, photometric and spectroscopic characteristics of the 52 \INSPIRE\ UCMGs, as well as each single-arm spectrum and the combined, restframed and rebinned ones are all publicly available from the ESO Science Portal\footnote{\url{https://archive.eso.org/scienceportal/home?data_collection=INSPIRE}}.

\section{Data analysis and stellar kinematics}
\label{sec:data}
The data reduction and analysis have already been extensively described in previous papers. We refer the readers to \citetalias{Spiniello+21} and \citetalias{DAgo23} for a complete description and for histograms of the distribution in Right Ascension (RA), Declination (DEC), redshift, as well as stellar masses, S\'ersic indices and axis ratios. 
In this final data release (DR3), we add 12 new systems to those already presented in DR1 and DR2, completing the spectroscopic analysis for the 52 galaxies in the \INSPIRE\ sample.

We highlight the main steps of the analysis in the following. 
As in the previous papers, we reduce the data with the standard ESO XSH pipeline (v3.5.3) under the ESO Reflex Workflow (\citealt{Freudling+13}, version 2.11.3) until the creation of a 2-dimensional (2D) spectral frame per each observation block and per each arm of the detector. Then, we extract a 1-dimensional (1D) integrated spectrum encapsulating 50\% of the light. We obtain in this way an integrated 1D spectrum comparable in light fraction to that extracted at the \Reff, but taking into account the seeing during observations.
Given the very small apparent sizes of these UCMGs, the spectra are completely seeing dominated. Hence, we extract the R50 1D spectra starting from the surface brightness profiles of the 2D spectra and then integrating them up to the aperture that encapsulates 50\% of the total light but contains a mix of light from inside and outside the real effective radius (we refer the readers to \citetalias{Spiniello+21} for a more detailed description on the 1D extraction). 
Finally, we correct the VIS and NIR arms for telluric absorption lines, using the code \textsc{molecfit}\footnote{\url{http://www.eso.org/sci/software/pipelines/skytools/molecfit}}  (\citealt{Smette15}, v.4.2). 

Once a 1D spectrum is obtained for each arm (combining together the different scientific exposures), we measure the redshift and the SNR of each galaxy independently from the three arms, using the entire spectral coverage\footnote{We use the IDL code \texttt{}{DER\_SNR} (\citealt{Stoehr08}, \url{https://www.stecf.org/software/ASTROsoft/DER_SNR}) which estimates the SNR directly from the flux, assuming that the noise is uncorrelated in wavelength bins spaced two pixels apart and that it is approximately Gaussian distributed.}. Then, we convolve all of them to the same final resolution (FWHM$_{\mathrm{fin}} = 2.51$\AA, at restframe) and join them together, producing the final spectrum that we use to compute the stellar velocity dispersion and stellar population parameters, via full spectrum fitting, using stellar templates with the same FWHM. 

Both single-arm spectra and the combined and smoothed ones are publicly released as part of the \INSPIRE\ ESO Phase 3 collection\footnote{\url{https://archive.eso.org/scienceportal/home?data_collection=INSPIRE}}. A detailed description of how the arms have been joined together and how the data have been smoothed at the final resolution is provided in the data release description, while the final UVB+VIS smoothed spectra are shown in Appendix~\ref{app:spectra}. 

We stress that, although spectra covering the entire XSH wavelength range (from the UVB to the NIR) are obtained and made available to the astronomical community, for the moment we limit our analysis and based our results only on the UVB+VIS. This is mainly because the SSP modelling is much better undertsood in this wavelength range, while more investigation is needed in the NIR, especially regarding Carbon-sensitive indices which are systematically underestimated by the current SSP models \citep{Eftekhari22}.

\begin{table}
\centering
\caption{Spectroscopic properties of the \INSPIRE\ sample: redshift, SNR in the three arms and their mean, and velocity dispersion obtained from the smoothed and combined 1D spectrum. Galaxies are listed in descending order of mean SNR per \AA\ and split into three groups, as described in the text. }
\label{tab:spec_cat}
\begin{tabular}{lcccccc}
\hline
\hline
  \multicolumn{1}{c}{ID KiDS} &
  \multicolumn{1}{c}{z} &
  \multicolumn{1}{c}{SNR} &
  \multicolumn{1}{c}{SNR} &
  \multicolumn{1}{c}{SNR} &
  \multicolumn{1}{c}{SNR} &
%  \multicolumn{1}{c|}{SNR$_{\mathrm{UVB}}$} &
%  \multicolumn{1}{c|}{SNR$_{\mathrm{VIS}}$} &
%  \multicolumn{1}{c|}{SNR$_{\mathrm{NIR}}$} &
  \multicolumn{1}{c}{$\sigma_{\star}$} \\
  \multicolumn{1}{c}{ } &
  \multicolumn{1}{c}{} &
%  \multicolumn{1}{c|}{($\ \AA$)} &
  \multicolumn{1}{c}{UVB} &
  \multicolumn{1}{c}{VIS} &
  \multicolumn{1}{c}{NIR} &
  \multicolumn{1}{c}{mean} &
  \multicolumn{1}{c}{(km/s)} \\
\hline
  J1142+0012 & 0.1077 & 57.9 & 124.1& 58.2 & 80.0 & $129\pm6 $\\
  J1026+0033 & 0.1743 & 38.9 & 113.6& 56.9 & 69.8 & $225\pm11$\\
  J1417+0106 & 0.1794 & 39.9 & 107.7& 26.6 & 58.1 & $203\pm10$\\
  J2312-3438 & 0.3665 & 32.1 & 72.4 & 33.6 & 46.1 & $221\pm11$\\
  J0909+0147 & 0.2151 & 20.7 & 75.3 & 37.6 & 44.5 & $401\pm20$\\
  J0847+0112 & 0.1764 & 24.6 & 77.5 & 31.2 & 44.4 & $244\pm12$\\
  J0224-3143 & 0.3839 & 20.9 & 71.2 & 36.3 & 42.8 & $283\pm14$\\
  J1228-0153 & 0.2973 & 23.2 & 70.1 & 34.0 & 42.4 & $191\pm10$\\
  J1128-0153 & 0.2217 & 21.1 & 69.2 & 30.8 & 40.4 & $192\pm10$\\
  J1411+0233 & 0.3598 & 24.1 & 73.2 & 22.4 & 39.9 & $217\pm11$\\
  J0838+0052 & 0.2702 & 22.6 & 65.2 & 31.8 & 39.9 & $189\pm9 $\\
  J0918+0122 & 0.3731 & 17.6 & 70.2 & 31.6 & 39.8 & $242\pm12$\\
  J2327-3312 & 0.4065 & 19.6 & 72.8 & 25.8 & 39.4 & $227\pm11$\\
  J0321-3213 & 0.2947 & 21.9 & 66.4 & 26.7 & 38.3 & $211\pm11$\\
  J1436+0007 & 0.2210 & 21.1 & 67.2 & 24.2 & 37.5 & $193\pm19$\\
  J1447-0149 & 0.2074 & 24.7 & 64.7 & 22.1 & 37.2 & $187\pm9 $\\
  J1156-0023 & 0.2552 & 22.6 & 60.9 & 26.5 & 36.7 & $177\pm18$\\
  J0314-3215 & 0.2874 & 20.7 & 54.6 & 31.2 & 35.5 & $178\pm9 $\\
\hline
  J0326-3303 & 0.2970 & 21.4 & 54.5 & 26.7 & 34.2 & $173\pm17$\\
  J0226-3158 & 0.2355 & 22.9 & 58.7 & 20.5 & 34.0 & $185\pm19$\\
  J0240-3141 & 0.2789 & 17.9 & 54.5 & 26.9 & 33.1 & $216\pm22$\\
  J0920+0126 & 0.3117 & 17.9 & 55.6 & 24.7 & 32.7 & $190\pm19$\\
  J1438-0127 & 0.2861 & 17.9 & 59.4 & 20.2 & 32.5 & $218\pm22$\\
  J1114+0039 & 0.3004 & 19.5 & 54.0 & 23.7 & 32.4 & $181\pm18$\\
  J0920+0212 & 0.2800 & 17.0 & 55.0 & 21.9 & 31.3 & $246\pm25$\\
  J2204-3112 & 0.2581 & 14.4 & 54.1 & 20.4 & 29.6 & $227\pm23$\\
  J0317-2957 & 0.2611 & 20.1 & 51.7 & 15.5 & 29.1 & $187\pm19$\\
  J2359-3320 & 0.2888 & 15.6 & 49.1 & 21.2 & 28.6 & $267\pm27$\\
  J0917-0123 & 0.3602 & 12.2 & 50.3 & 22.9 & 28.5 & $239\pm24$\\
  J0316-2953 & 0.3596 & 14.3 & 46.4 & 24.4 & 28.3 & $192\pm19$\\
  J1040+0056 & 0.2716 & 11.5 & 46.7 & 26.2 & 28.1 & $240\pm24$\\
  J2305-3436 & 0.2978 & 14.3 & 46.8 & 21.1 & 27.4 & $295\pm30$\\
  J0211-3155 & 0.3012 & 13.9 & 46.7 & 21.4 & 27.3 & $245\pm25$\\
  J1202+0251 & 0.3298 & 14.7 & 45.9 & 21.1 & 27.3 & $165\pm17$\\
  J0844+0148 & 0.2837 & 12.9 & 45.0 & 23.3 & 27.1 & $224\pm22$\\
  J0842+0059 & 0.2959 & 12.4 & 41.5 & 25.5 & 26.5 & $324\pm32$\\
  J1154-0016 & 0.3356 & 16.6 & 42.8 & 19.6 & 26.3 & $163\pm16$\\
  J0904-0018 & 0.2989 & 12.6 & 44.3 & 21.4 & 26.1 & $205\pm21$\\
  J2257-3306 & 0.2575 & 17.8 & 40.0 & 20.0 & 25.9 & $185\pm19$\\
  J2202-3101 & 0.3185 & 13.1 & 47.6 & 17.0 & 25.9 & $221\pm22$\\
  J0857-0108 & 0.2694 & 15.7 & 43.3 & 18.6 & 25.8 & $166\pm17$\\
  J1218+0232 & 0.3080 & 14.6 & 42.0 & 18.9 & 25.2 & $171\pm17$\\
 \hline
  J1420-0035 & 0.2482 & 13.4 & 39.8 & 15.5 & 22.9 & $209\pm31$\\
  J1449-0138 & 0.2655 & 10.2 & 40.3 & 15.9 & 22.1 & $192\pm29$\\
  J1456+0020 & 0.2738 & 11.9 & 39.6 & 14.4 & 22.0 & $194\pm29$\\
  J1414+0004 & 0.3030 & 11.7 & 37.4 & 15.3 & 21.5 & $205\pm31$\\
  J2356-3332 & 0.3389 & 11.5 & 34.2 & 14.8 & 20.1 & $162\pm24$\\
  J1457-0140 & 0.3371 & 12.5 & 34.1 & 12.4 & 19.7 & $203\pm30$\\
  J1402+0117 & 0.2538 & 12.4 & 34.0 & 11.9 & 19.5 & $166\pm25$\\
  J1412-0020 & 0.2783 & 10.5 & 31.1 & 14.2 & 18.6 & $339\pm51$\\
  J1527-0012 & 0.4000 & 7.1  & 32.7 & 14.6 & 18.1 & $237\pm36$\\
  J1527-0023 & 0.3499 & 9.0  & 30.4 & 13.5 & 17.7 & $188\pm28$\\
\hline
\hline
\end{tabular}
\end{table}

We re-derive stellar velocity dispersion values ($\sigma_{\star}$) for all the \INSPIRE\, objects, using the same parameters in the Penalised Pixel-fitting software\footnote{\url{https://pypi.org/project/ppxf/}} (\ppxf; \citealt{Cappellari04,Cappellari17}) for each of the galaxies. We fix the wavelength range of the fit to [3500-9000]\AA, and the additive Legendre polynomial degree (DEGREE) to 13. We use the keyword CLEAN to perform a sigma-clipping on the spectra and clean them from residual bad pixels. As stellar library for the fit, we use the E-MILES Single Stellar Population (SSP) models \citep{Vazdekis16}.  
These models are an extention of the MILES models \citep{Sanchez-Blazquez+06, Vazdekis15}, 
covering the spectral range [1680-50000]\AA\  and have a spectral resolution of FWHM$\sim2.5$\AA\ in the wavelenght range used for the fit. 

In \citetalias{DAgo23} we have performed a comprehensive series of tests to assess the uncertainties on the velocity dispersion measurements arising from changing the fitted region, the mask for bad pixels, the degree of the additive polynomial, the resolution and SNR of the input spectrum, and the number of moments of the line-of-sight velocity distribution fitted by the routine. By means of a bootstrapping procedure realised ad-hoc for the purpose, we showed that the uncertainties on $\sigma_{\star}$ are of the order of 10-15\% for the lowest SNR spectra ($\sim20$ per \AA) and $\sim5$\% for the highest SNR ones ($\sim80$ per \AA), as visible from Figure~6 in the  \citetalias{DAgo23}. 
Hence, we split the 52 UCMGs into three groups and assign a different uncertainty ($\Delta\sigma_{\star}$) to them. The low SNR group, with a relative error of 15\% on the velocity dispersion, comprises spectra with mean SNR ranging from 15 to 25. The objects in the medium SNR group, with spectra with mean SNR from 25 to 35, have an uncertainty on the stellar velocity dispersion of 10\%, while objects in the high SNR group ($>35$) have an uncertainty of 5\%. 

Table~\ref{tab:spec_cat} presents the  characteristics of the \INSPIRE\ sample that have been estimated from the 1D spectra directly. The typical uncertainties on the redshift are of the order of 0.1-0.5\%. 
The last column of the table reports the stellar velocity dispersion values obtained via full spectral fitting, as described above, and the associated uncertainties. The objects are ordered from the highest to the lowest mean SNR, and the three groups are separated by a horizontal line. We note that we take the arithmetic mean of the SNRs computed from each single arm (listed in the table), using all the pixels, to split the systems into the groups. 

\begin{figure}
    \centering  \includegraphics[width=\columnwidth]{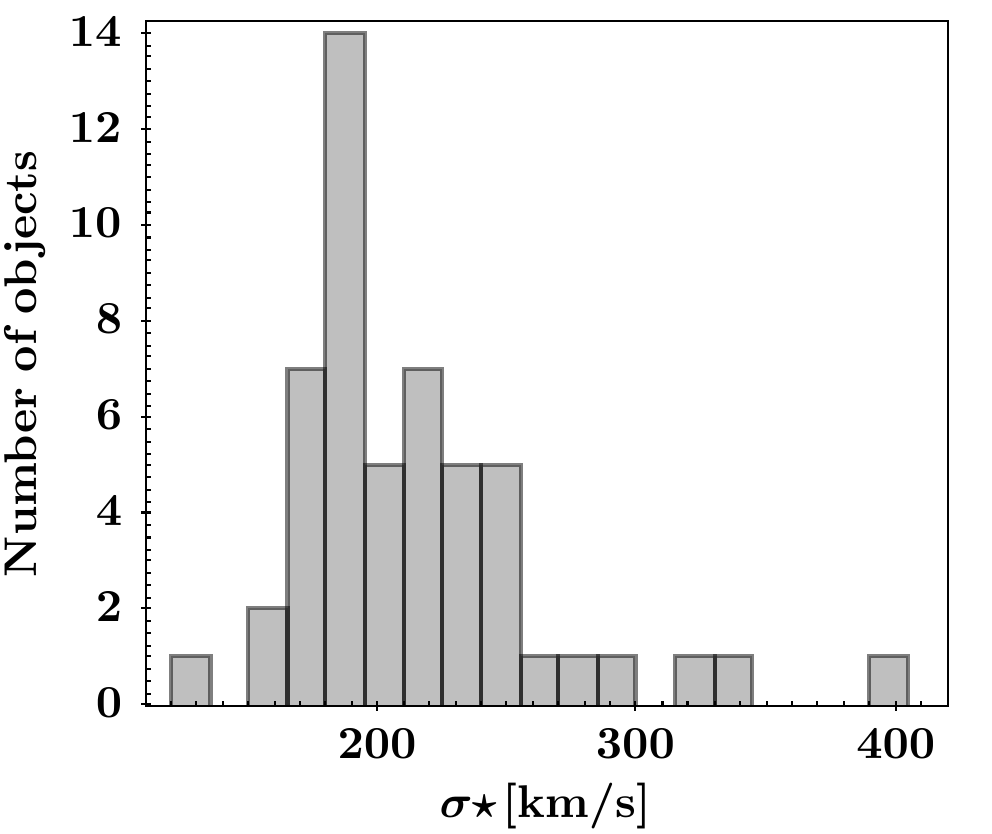}
    \caption{Distribution of the stellar velocity dispersion measurements for the entire INSPIRE sample. }
    \label{fig:vdisp_histo}
\end{figure}

Figure~\ref{fig:vdisp_histo} shows the distribution of $\sigma_{\star}$ for the final INSPIRE sample. As expected from the range in stellar masses covered by the galaxies, the peak in velocity dispersion is $\sim215$ \kms, with however a long tail towards larger velocity dispersion values. In Section~\ref{sec:results}, we will show that this tail correlates with the DoR of the objects, as already hinted in \citetalias{Spiniello+21}: galaxies with a higher DoR have overall larger stellar velocity dispersion with respect to galaxies of equal stellar mass but lower DoR.

\section{Stellar populations analysis}
\label{sec:stel_pop}
Following the steps of previous \INSPIRE\ papers, we use 
line-indices to derive the [Mg/Fe] abundance for each UCMG, and then the Penalised Pixel-fitting software (\ppxf; \citealt{Cappellari04,Cappellari17}) to perform a full spectral fit of the UCMGs spectra,  computing integrated, mass-weighted ages and metallicities. 
The two processes are described in the next sections. 

For both the line-indices and the full-spectral fitting, we use the MILES models described in \citet{Vazdekis15} 
with BaSTI theoretical isochrones\footnote{
\href{http://www.oa-teramo.inaf.it/BASTI}{http://www.oa-teramo.inaf.it/BASTI}.} \citep{Pietrinferni04, Pietrinferni06}. These models allow the user to investigate a broad range of stellar ages from 30 Myr to 14.5 Gyr (sampled at logarithmic steps), metallicities in the range $-1.79<$[M/H]$<+0.40$ and a suite of stellar Initial Mass Function (IMF) slopes. The fitted range for the \ppxf\ stellar population analysis is [3525-7500]\AA, which  corresponds to entire range covered by the SSP libraries. 

%\footnote{Given the sampling of the SSPs in age ($\Delta_{\text{age}} = 0.5$ Gyr), we choose, for each galaxy, the model with an age as close as possible to the age of the Universe at that redshift.}, with steps of $\Delta t = 0.5$ Gyr, and five %seven 

%When constraining the stellar population parameters from integrated spectra of galaxies, a wide wavelength range is necessary in order to break the age-metallicity degeneracy  \citep{Worthey+94} and to decouple the variation of the chemical elements from variation in other parameters.  Hence, we use as input spectra the combined UVB+VIS+NIR, covering the restframe wavelength range [3000-10000]\AA\, at a resolution of FWHM = 2.51\AA\, identical to that of the single stellar population (SSP) models, which minimises template-mismatch. 

\begin{figure}
    \centering    \includegraphics[width=\columnwidth]{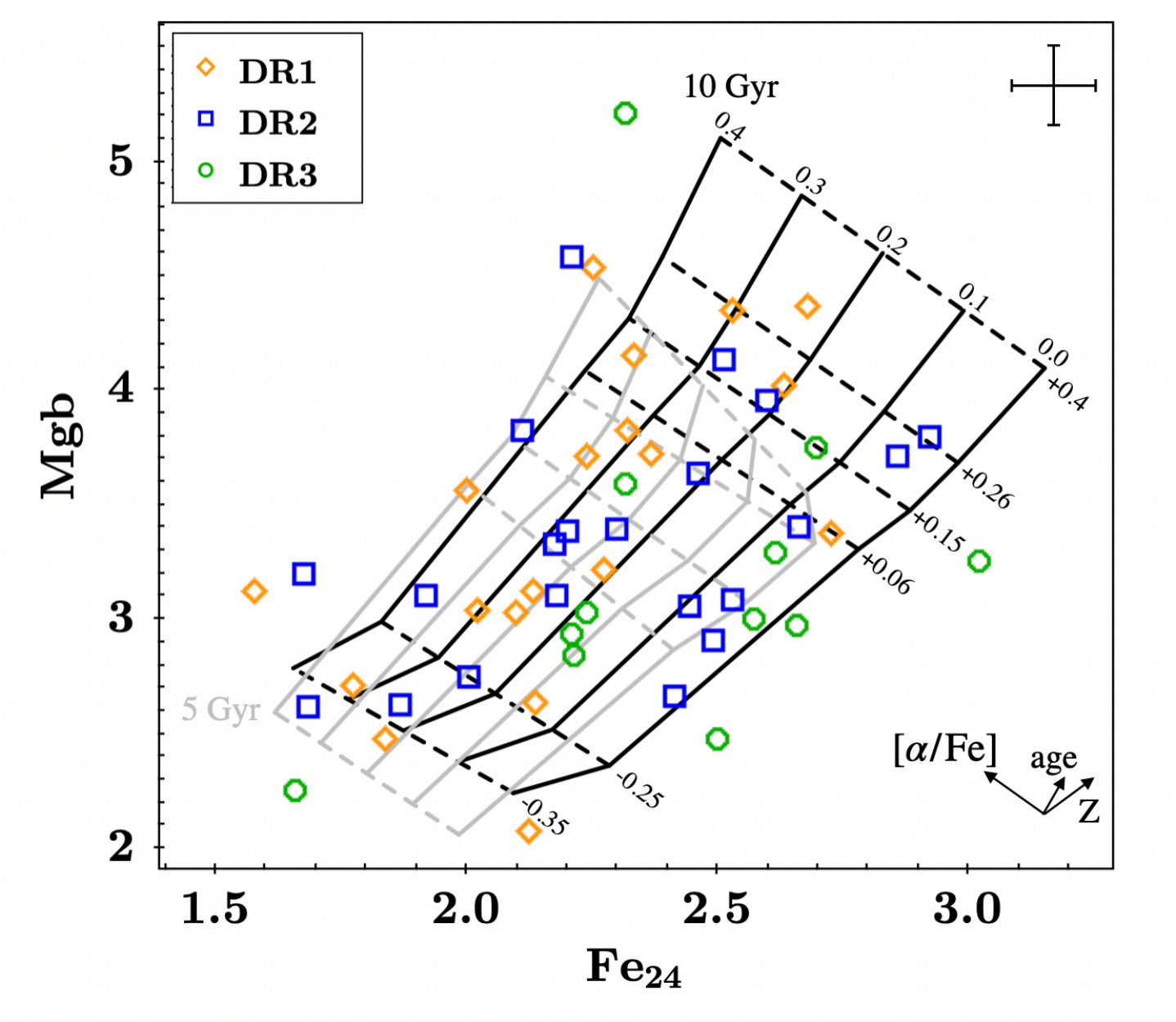}
    \caption{Mg$_b$-Fe24 index--index plot used to infer the [Mg/Fe] for all the INSPIRE objects. The grid shows MILES SSPs with two different ages (10 Gyr in black and 5 Gyr in grey), covering a range of metallicities  and \afe\ values. The direction of variation for age, metallicity and \afe\ is given by the arrows in the bottom-right corner. 
    The \afe\ is orthogonal to the other parameters and hence can be  inferred even without a precise estimate of age and metallicity.  The \INSPIRE\ galaxies are colour-coded and have different symbols, according to the Data Release to which they belong to. A typical error-bar is shown on the right-top corner of the plot. } %Line strengths for both galaxies and SSPs are calculated after convolving the spectra to a common resolution of $\sigma = 402$ \kms. }
    \label{fig:alphafe}
\end{figure}

\begin{figure*}
    \centering  \includegraphics[width=\textwidth]{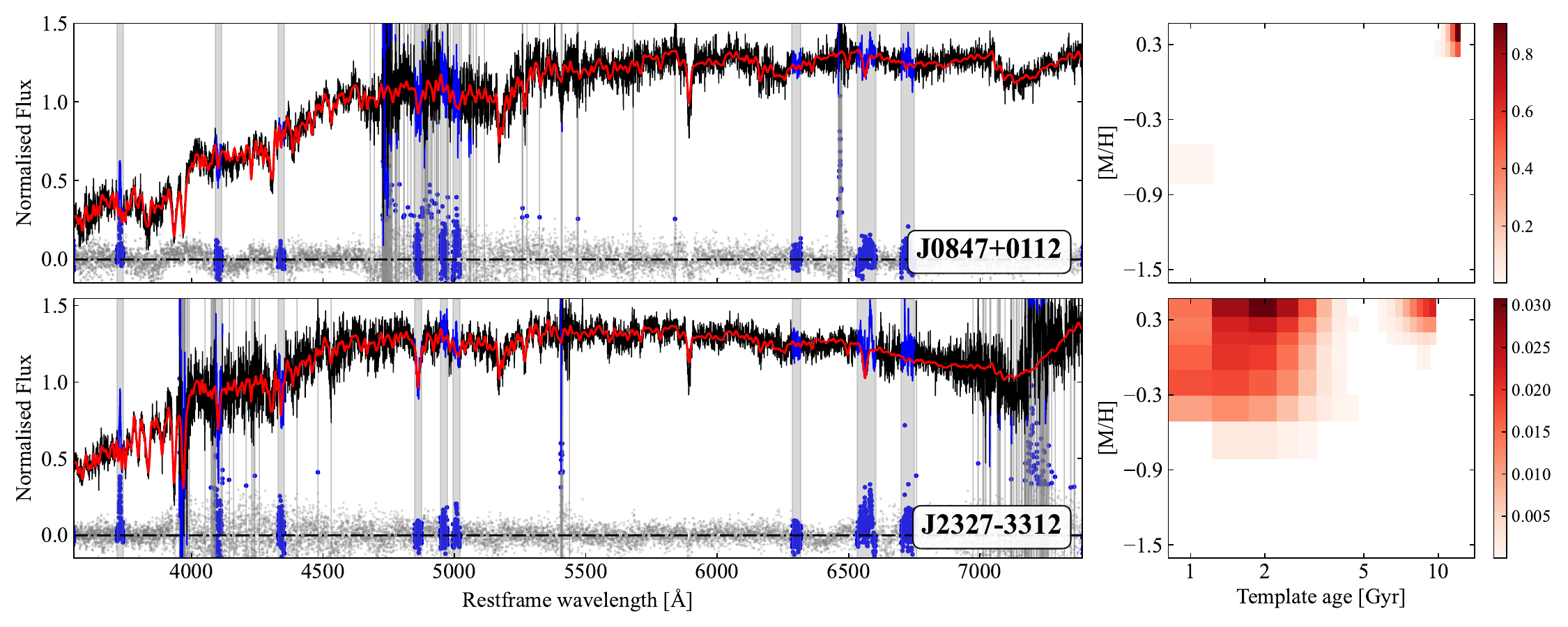}
    \caption{The \ppxf\ fits for the oldest (top) and the youngest (bottom) galaxies. The left panels show the fits, with the galaxy spectrum in black, the SSP best-fit template in red, residuals in grey, and pixels masked out from the fit in blue. The right panels show the age-metallicity density map obtained for the maximum regularized case. They represent the mass-weighted fraction assigned by \ppxf\ to each input SSP.}
    \label{fig:ppxf_res}
\end{figure*}
 
\subsection{SSP-equivalent [Mg/Fe] abundances from line-indices}
The [Mg/Fe] abundance is often used as a proxy for the time-scale efficiency of the star formation episode. For a fixed IMF, a large [Mg/Fe] value indicates a very quick quenching of star formation, which occurs before Type Ia supernova explosions can pollute the interstellar medium with iron \citep[e.g., ][]{Matteucci94, Thomas+05, Gallazzi+06, Gallazzi21}. 
Hence, since relics should have formed the (almost) totality of their stellar mass in a very fast star formation burst, they are expected to have a very large [Mg/Fe] abundance. 
Indeed, a high, super-solar ratio has been measured for the three local relics \citep{Yildirim17, Ferre-Mateu+17, Martin-Navarro18} and for the confirmed relics in \citetalias{Spiniello+21}. 

Since measuring \afe\ from full-spectral fitting requires further tests (see e.g., \citealt{Vazdekis15} and \citealt{Liu20}), 
%Since the original version of the \ppxf\, code only works in the age-metallicity parameter space, 
we estimate the [Mg/Fe] via the analysis of line-index strengths from Mg and Fe lines. To reproduce exactly what has been done in \citetalias{Spiniello+21}, we use the Mg$_b$ (5177\AA) and the mean of  24 different iron lines present in the spectral range [$3500-7000$]\AA. Combining many different Fe-lines 
helps to minimise the dependency on other elemental abundances (e.g. [Ti/Fe]) which might fall in the Fe-indices bandpasses. We have however tested our result against a different combination of Fe-lines, finding results overall consistent within the uncertainties. 

We use the code SPINDEX from \citet{Trager08} to calculate the strengths of these features from the galaxies' spectra, all convolved to a common velocity dispersion of 402 \kms, equal to the maximum stellar velocity dispersion (the same of  J0909+0147). 
We then measure the line-index strengths of the same spectral features in SSP models brought to the same resolution. We use two sets of ages (5 and 10 Gyr, justified by the results of DR1), metallicities ranging from $-0.35$ to $+0.4$, and a bimodal IMF with $\Gamma=1.3$ (resembling a Kroupa IMF).  
We note that the MILES models do not allow to control the [Mg/Fe] directly, but this can be approximated to \afe, which is the parameter we can change in the SSPs. 
The version of the MILES models we employ here is available only at two different \afe\ abundances: 0.0 (solar) and 0.4. We note that a new version of the MILES models has very recently been released \citep{Knowles23}. These models allow for variation in \afe\ from $-0.2$ to $+0.6$ with steps of $0.2$. However, as pointed out by the authors of the SSP models, for \afe\ enhancements from scaled-solar to 0.4, the abundance pattern predictions from the new models and that obtained from the old ones show only minimal differences for old SSP ages over a wide range of total metallicities.

The result of the line-indices analysis can be visualised via the Mg$_b$-Fe$_{24}$ index--index plot, shown in Figure~\ref{fig:alphafe}. In this plot, age and metallicity are degenerate, both increasing from the bottom left corner to the top right one. The \afe\ varies instead in an orthogonal direction, as shown by the arrows in the bottom-right corner of the figure. Therefore, this particular pair of indices allows us to estimate the [Mg/Fe] for all the UCMGs, even before having a precise constraint on the age and metallicity. In order to build a finer grid, we  linearly interpolate the models (0.0 and 0.4 in \afe) in steps of $\Delta$\afe=0.1. We prefer not to extrapolate the models outside the original boundaries. 
%In Figure~\ref{fig:alphafe}, the \INSPIRE\ galaxies are shown as coloured symbols, which are changing in shape and colour according to the correspondent Data Releases, as highlighted in the legend.  
We then directly estimate the [Mg/Fe] from the plot, taking as value the \afe\ of the closest model to each point on the 10 Gyr grid. Given that there is no prior on age and metallicity and these will slightly shift the model grids, we assign an uncertainly of 0.1 to all the [Mg/Fe] values, which is equal to the step between the models. We could have computed a more precise estimate of the [Mg/Fe] by interpolating the models over a fine grid, also in age and metallicity. However we decided not to do so since these ``SSP-equivalent" [Mg/Fe] estimates, which are listed in the second column of  Table~\ref{tab:stelpop}, will only be used to select SSP models with a given \afe\ as input in the full-spectral fitting, as detailed in Section~\ref{sec:stel_pop_ppxf}. 
We note that for 6 DR1 systems we find an estimate that deviated by $\pm0.1$ from that estimated in \citetalias{Spiniello+21}. This might be due to small differences in the line-indices measurements that could be caused by the different smoothing applied here\footnote{In DR1 the largest velocity dispersion value was $\sigma=300$ \kms, and hence both galaxies and SSP were convolved to that resolution.}.
 
\subsection{Mass-weighted stellar ages and metallicities from full-spectral fitting}
\label{sec:stel_pop_ppxf}
Mass-weighted stellar ages and metallicities are obtained with \ppxf\, performing a  full-spectral fitting on the integrated 1D R50 spectra. 
Also in this case, for consistency, we use the MILES single SSP models as stellar templates for the fit. 
For our purposes, we use models with ages from 1 Gyr to %12 Gyr 
the age of the Universe at the redshift of each object\footnote{Given the sampling of the SSPs in age ($\Delta_{\text{age}} = 0.5$ Gyr), we allow models up to 0.5 older than the age of the Universe at the galaxies' redshifts.} 
(logarithmic sampled with steps of $\Delta \log t$ = 0.5 Gyr). 
In future \INSPIRE\ publications, we will extend the analysis to the UVB arm (at wavelengths bluer than 3500\AA) in order to increase the sensitivity to younger ages. This is motivated by the finding of \citet{Salvador-Rusinol2021_nature} and \citet{Salvador-Rusinol22}, which detected residual sub-one per cent recent star formation in massive early-type galaxies and in the local relic NGC1277. 
%We explore ten different total metallicity bins (in terms of the total stellar metal abundance [M/H]), ranging from  $-1.49$ to $+0.40$, but keep the stellar IMF slope fix to a bimodal-distribution with a low-mass end slope of  $\Gamma$ = $1.3$. 
%Finally, as already done in DR1 and in the previous section, we use the publicly available models with \afe=0.0 and \afe=0.4, and linearly interpolate them to cover the \afe\ space with steps of 0.1. 

The \ppxf\ fit is performed in 2D, only letting age and metallicity vary in the models. In fact, as already done in \citetalias{Spiniello+21}, we use, for each galaxy, only models with the \afe\ ratio equal to that estimated from line-indices. In future \INSPIRE\ publications, we will improve on this aspect by implementing a \ppxf\ fit in 3D, or even 4D, also varying the IMF slope, which is kept fixed to a bimodal-distribution with a low-mass end slope of  $\Gamma$ = $1.3$. 
In fact, %for now, although 
the DR1 data hinted at a  difference in the low-mass end of the IMF slope between relics and non-relics \citep{Martin-Navarro+23}, however, a much higher SNR is necessary to measure IMF variations and break the degeneracies with the other stellar population parameters \citep[e.g.,][]{Conroy_vanDokkum12b, LaBarbera+13_SPIDERVIII_IMF, Spiniello+14}. 
%we do not change the IMF  and restrict the fit to wavelength bluer than 7500\AA, where the dependency of IMF is minimal \citep{Conroy13, Spiniello+14}. This is justified by the fact that in order to detect variation on its low-mass slope, a relatively higher SNR is necessary.
We will investigate how a variable IMF might influence the other stellar population parameters and whether or not the slope correlates with the DoR in a forthcoming paper, stacking galaxies with similar DoR or other parameters. 
%Finally,  the original models allow for only two [$\alpha$/Fe] values: 0.0 (solar) and 0.4. We therefore linearly interpolate the SSPs bulding models with [$\alpha$/Fe] = $\{0.0, 0.1, 0.2, 0.3, 0.4\}$.

Following  \citetalias{Spiniello+21}, we perform the \ppxf\, fit using the two most extreme values of the regularisation parameter, thus deriving the minimally (REGUL = 0, hereafter `unr') and maximally smoothed (REGUL=MAX\_REGUL, hereafter `rmax') solutions consistent with each  spectrum and its SNR. 
The MAX\_REGUL changes from system to system and it is calculated by rescaling the noise until the $\chi^2$ increases by $\Delta\chi^2 = \sqrt{2 \times \text{DoF}}$, where the Degree of Freedom (DoF) is approximated to be equal to the number of pixels used in the fit. 
Unlike for the kinematics, we only use multiplicative polynomials for the stellar populations to preserve the line strengths.  We set the multiplicative polynomial degree MDEGREE=10 
(see the tests performed in Appendix A of the \citetalias{Spiniello20_Pilot}).
%To estimate reliable uncertainties on the mass-weighted ages and metallicities, we run a bootstrap routine repeating each fit 256 times, randomly rescaling the RMS noise obtained during the first fit, but maintaining the reduced $\chi^{2}$ between 0.9 and 1.1. 
To estimate reliable uncertainties on the mass-weighted ages and metallicities, we run a bootstrap routine consisting in repeatedly fitting a \ppxf\ model to a spectrum obtained by adding random noise to the original one. The random noise is within 3 times the RMS obtained by subtracting from the original spectrum a median smoothed version of itself. We repeat this procedure 256 times and use the standard deviation as uncertainty on the measured stellar population parameters. At this stage, we also vary the MDEGREE (within $\pm 2$) and the REGUL parameter (within MAX\_REGUL$-5$ and MAX\_REGUL). 
%At this stage, we also vary the MDEGREE (within $\pm 4$) and the REGUL parameter (from 0 to the MAX\_REGUL). 

Figure~\ref{fig:ppxf_res} shows two examples of \ppxf\ fits for the olderst (top) and the youngest (bottom) UCMGs. In the left panels, the galaxy spectrum is plotted in black, while the best-fit SSP template is overplotted in red. The right panels show the mass and metallicity inferences from the weights of the SSP models, used to build the best-fit, in the rmax case. 

\begin{figure}
    \centering  \includegraphics[width=\columnwidth]{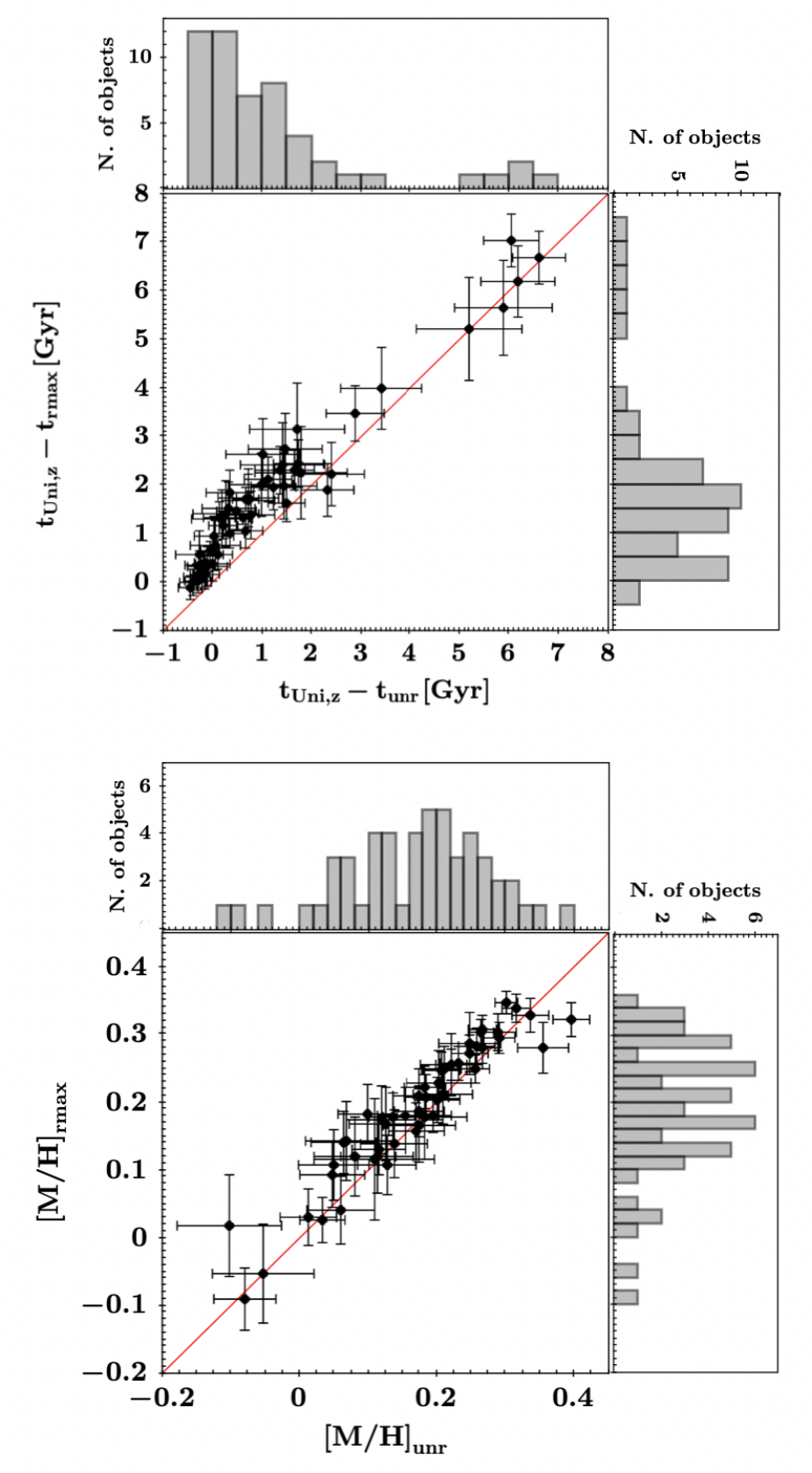}
    \caption{Scatter plots of the formation ages (top) and stellar metallicities (bottom) derived from the unr versus the rmax \ppxf\ fits. As expected, ages are younger in the latter case. The points on the  metallicity plot (lower panel) lie on the one-to-one line, shown in red. The histograms show the distribution of the parameters in the two cases separately. }
    \label{fig:stel_pop_histo}
\end{figure}

Results obtained in the two extreme \ppxf\ configurations (unr and rmax) can be seen in the scatter plots of Figure~\ref{fig:stel_pop_histo}. 
In the top panel, we plot the formation epoch of each galaxy, defined as the difference between the age of the Universe at the redshift of each object and the stellar age derived from full spectral fitting $(t_{\text{Uni,z}} - t_{\text{gal}})$ for the unr case ($x$-axis) versus the rmax one ($y$-axis). This quantity is a proxy for the formation age of the systems: the closer this number is to zero, the earlier the object assembled
its stellar mass in the history of the Universe. Smoother solutions clearly imply younger ages. 
Since in principle, regularized and unregularized solutions should be statistically equivalent, the fact that they produce mass-weighted mean ages that are only marginally consistent within the plotted error bars, demonstrates that our bootstrapping routine produces uncertainties that only take into account the uncorrelated noise in the spectra but are not sensitive to  systematics arising from the regularisation of the fits.

The bottom scatter plot shows the different estimates of [M/H] in the two cases. As expected, younger ages (in the rmax case) translate into higher metallicities. 
However, since the difference between the two \ppxf\ runs is small, the regularisation does not play a major role in constraining the stellar metallicity.

The histograms on the corners of each panels show the distribution of formation ages and metallicities for the two regularisations.

%Examples of the \ppxf\ fit. Galaxies are ordered from the oldest (top) to the youngest (bottom). The left panels show the fits, with the galaxy spectrum in black, the SSP best-fit template in red, residuals in grey, and pixels masked out from the fit in blue. The middle panels show the age-metallicity density plot obtained for the rmax case. In the right panels, we plot the fraction of stellar mass assembled from the Big Bang to the age of the Universe at the system's redshift. Solid black line shows the unr fit, green  the rmax one. The grey zone indicates how the (unr) SFH changes using SSP models with \afe=[Mg/Fe]$\pm0.1$. }    

All stellar population results are summarised in Table~\ref{tab:stelpop}, where we report the [Mg/Fe] values estimated from line-indices analysis and the mass-weighted ages and metallicities for the unr and the rmax cases. 
These quantities are used to compute SFHs, and in particular the fraction of stellar mass assembled in cosmic time. 
Specifically, this latter quantity is computed from the age-[M/H] 'weights fractions maps' (density maps) shown in the right panels of Figure~\ref{fig:ppxf_res}, by flipping the age axis and marginalising over metallicity. 
This quantity will be used in the next section to estimate the DoR and assess how many of the \INSPIRE\ UCMGs can be classified as relics of the ancient Universe.

\section{Results}
\label{sec:results}
In \citetalias{Spiniello+21} we have computed the fraction of mass assembled within 3 Gyr after the Big Bang (\Mfrac), i.e. by $z=2$, assumed to be the end of the first phase \citep{Zolotov15}, the cosmic time at which 75\% of the stellar mass was in place ($t_{75}$) and the time of final assembly, i.e the cosmic time at which the entire totality (100\%) of the stellar mass was in place ($t_{\rm fin}$). This allowed us to weigh the contribution by ancient stars separating them from those formed in later episodes of star formation (if any) and how much they extended in cosmic time. 
Here we compute the same quantities for all the 52 \INSPIRE\ objects and list them in Table~\ref{tab:stelpop}, alongside with the stellar population results. 

To produce the stellar mass assembled in cosmic time, we start from the density maps, like the ones shown in the right panels of Figure.~\ref{fig:ppxf_res}. We flip the age axis and sum over all metallicity values, computing in this way the fraction of mass assembled at each age bin since the Big Bang to the redshift of the galaxy. 
In \citetalias{Spiniello+21} we use the  \Mfrac\ and $t_{75}$to classify UCMGs in relics and non-relics. In particular, the quantities we quote in the table are the 
%In particular, the first column of the third block list the fraction of stellar masses assembled within 3 Gyr after the Big Bang (\Mfrac), i.e. by $z=2$, assumed to be the end of the first formation phase by \citet{Zolotov15}). 
%These numbers are computed as
minimum \Mfrac\ and the maximum $t_{75}$ of the 4 different estimates  obtained from four different \ppxf\ configurations: \\
i) an unregularised fit using models with \afe\ equals to the [Mg/Fe] value estimated from line-indices,\\
ii) the fit using the maximum regularisation and models with \afe=[Mg/Fe],\\
iii) an unregularised fit using models with  \afe=[Mg/Fe]$+0.1$,\\
iv) an unregularised fit using models with \afe=[Mg/Fe]$-0.1$.  \\
The uncertainty is assumed to be the standard deviation between the same four estimates. 

The uncertainties we associate with \Mfrac\ and $t_{75}$ are assumed to be the standard deviation on the four measurements. 
It is very interesting to note that high \Mfrac\ have much smaller uncertainties. This is because for extremely peaked SFHs, the \ppxf\ configurations do not play a major role and the four estimates are very similar. This is not true for systems with a more time-extended SFH. 
Finally, the remaining two columns of the same block list instead the final assembly time ($t_{\text{fin}}$, 100\% of the mass formed) and the age of the Universe at the redshifts of the objects (t$_{\mathrm{Uni, z}}$). 

%The minimum mass has been used in \citetalias{Spiniello+21} to classify galaxies as relics. The mean mass fraction is useful to constrain the possible, averaged star formation history at high z, given the uncertainties of the methodology. 
%The objects are ordered from the highest  minimum \Mfrac\ to the lowest. For galaxies with equal values, the one with the highest mean \Mfrac\ is listed before. 

\subsection{The degree of relicness}
In \citetalias{Spiniello+21}, we used the minimum fraction  (among those obtained from different \ppxf\ setups) of stellar mass assembled at $z=2$ and the maximum cosmic time at which 75\% of the stellar mass was in place
to classify the 19 systems analysed there into three groups: extreme relics, relics and non-relics. 
%assembled by the end of the first phase of the two-phase formation scenario %and the final time of assembly (time at which the entire totality of the stellar mass was in place) 
%In \citetalias{Spiniello+21} we have defined relics as systems that had formed more than 75\% of the stellar mass by $z=2$, or equivalenty, systems for which  $t_{75} \le 3$ Gyrs. 
%from the minimum (among these obtained from different \ppxf\ setups) mass assembled at $z=2$ and the cosmic time at which 75\% of the stellar mass was in place. 
Here, we also consider the cosmic time of final assembly in order to resolve the diversity of the entire \INSPIRE\ sample. We need to be able to disentangle between an object that had formed the majority of its stellar mass early on in cosmic time but then underwent a very time-extended SF from an object that assembled a similar quantity of stellar mass during the first burst, but then completed its formation still at high-z.

In addition, this time, we combine these three quantities together defining a number for each object, hence expressing the relicness with a simple parameter that quantifies the mean SFH and that gives a sense of the diversity of galaxies in terms of their stellar populations. The degree of relicness (DoR) is defined as: 

\begin{equation}
\text{DoR}=\left[ 
f_{M^{\star}_{t\text{BB}=3}} + \frac{0.5 \text{Gyr}}{t_{75}} + \frac{0.7 \text{Gyr} +(t_{\text{Uni}}- t_{\text{fin}})}{t_{\text{Uni}}} \right] \times \frac{1}{3}
\end{equation}

%\textcolor{red}{TO BE DISCUSSED!SHOULD WE USE TFIN OR T75 here? OR A COMBINATION OF THEM? Plots are with t75, but table is still with tfin!DO WE NEED TO'RELATE'TO DR1?}

%where the multiplying factor (0.582) is used to rescale the DoR in order to have it
The equation is motivated by the very recent findings of \citet{Labbe23} that it takes about 500-700 Myrs to form a galaxy with $M_{\star}\sim10^{11}M_{\odot}$. Hence, the minimum time to form 75\% of the stellar mass is set to 0.5 Gyr, while the minimum time to assemble all the mass is set to 0.7 Gyr. The third addendum is related to the final assembly time (i.e. the time at which a galaxy has formed the 100\% of its stellar mass), but rescaled to the age of the Universe at the redshift of each galaxy. 

The average of these three single addenda in the equation is allowed to 
vary from 0 to 1, with 1 denoting systems that have formed the totality of their stellar mass in the shortest possible time\footnote{As a reference, NGC1277, the most extreme relic known so far in the local Universe has a DoR of $\sim0.95$}.  
A higher DoR indicates an earlier formation epoch with almost no contribution from stars brought in through accretion or formed in later SF episodes. 
A lower DoR instead means that, although a fraction of stars are old and were formed at very high-z, there is a non-negligible percentage ($>25\%$, according to the DR1 threshold) of later-formed populations with different ages and metallicities.  Hence galaxies with low DoR are characterised by a much more time-extended star formation history. 
However, it is important to stress that the definition of the DoR, linearly combining the \Mfrac, the $t_{75}$ (middle) and the $(t_{\rm Uni}-t_{\rm fin})/t_{\rm Uni}$ is motivated only by the stellar population analysis and the choices made in \citetalias{Spiniello+21}. 

\begin{figure}
%    \centering  
\includegraphics[width=\columnwidth]{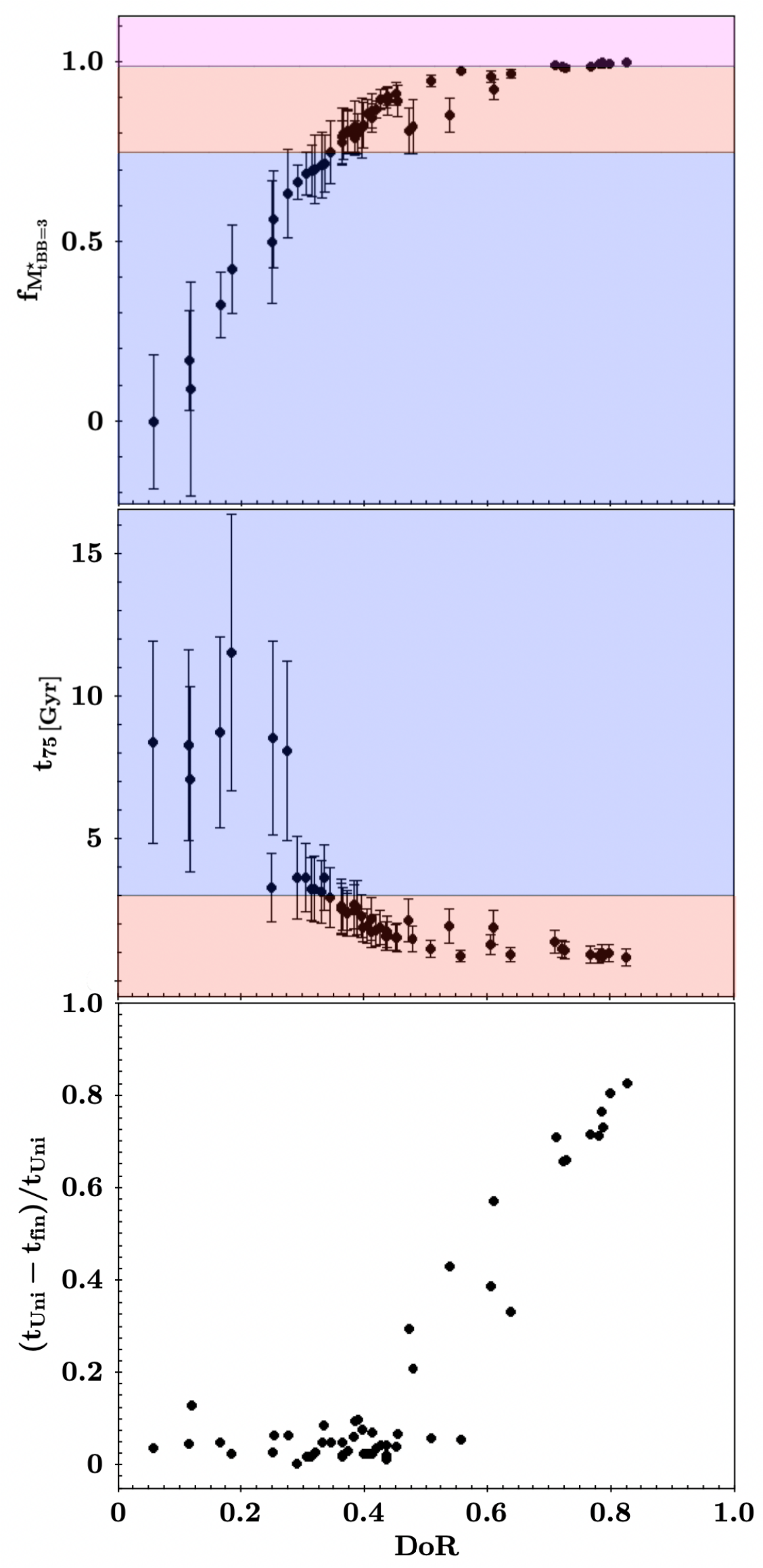}
    \caption{The DoR versus the fraction of mass formed up to 3 Gyr after the Big Bang (top), the cosmic time at which the galaxy have formed 75\% of their stellar mass (middle) and the time of final assembly (bottom). Horizontal lines show the thresholds corresponding to extreme relics (purple zone, upper plot) and relics (blue and red regions, first two plots) used in DR1.  }
    \label{fig:dor}
\end{figure}

We caution the reader that a one-to-one comparison between the relic definition in \citetalias{Spiniello+21} and this updated one is not straightforward. In fact, in DR1, we analysed two 1D extracted spectra for each galaxy (R50 and optimalli extracted, see \citealt{Spiniello+21} for more details) and considered the mean (lower limits for the) fraction of mass and $t_{75}$ when classify UCMGs in relics and non-relics. Nevertheless, a direct comparison between the ages and metallicities obtained in the two DRs is presented in Appendix~\ref{app:comparison}. Overall, a very good agreement is found, with only two galaxies presenting different stellar population parameters: J0226-3158, which is $\sim4$ Gyr younger now, and J0316-2953 for which the metallicity is lower in this DR3 with respect to the super-solar value inferred in DR1. J0226-3158 is the only object for which the relic classification has dramatically changed in this DR3. In DR1 we inferred a mean \Mfrac of 0.685, while now the fraction went down to $0.09\pm0.3$. For the 4 extreme relics already presented in DR1, nothing has changed here: they have still formed the great majority of their stars (\Mfrac$>0.99$) at $z>2$. 

In a forthcoming publication, already in preparation, we will use a more sophisticated machine-learning based grouping algorithm to identify `families' within the 52 UCMGs. We will feed the code with the morphological, photometrical, kinematical and stellar population parameters searching for clustering and grouping, and then comparing these results with the DoR grading.

Figure~\ref{fig:dor} shows the DoR defined above plotted versus the \Mfrac\ (top), the $t_{75}$ (middle) and the 
$(t_{\rm Uni}-t_{\rm fin})/t_{\rm Uni}$ (bottom). The coloured regions in the first two panels identify the zone where non-relics (blue), relics (red), and extreme relics (magenta) will lie, following the \citetalias{Spiniello+21} approach. The same colour code as in DR1 is used in these panels. 
However, from the figure, it appears that the definition used in DR1 does not allow one to really appreciate the diversity of the 52 UCMGs, spanning different final times and fractions of masses formed at high-$z$. 
For instance, the objects that would not pass the threshold of \Mfrac$\ge0.75$, cover a very large range of fraction of mass formed at high-z, from $0.72\pm0.09$ to as little as $0.00\pm0.18$. 
Also, some of them have $t_{75}$ only slightly larger than 3 Gyr while some others have much larger values.  This is also valid for the final assembly time and objects that have been classified as relics in DR1. 
In conclusion, the DoR is a much robust and informative way to quantify the fraction of mass formed at early times and the SFHs for all the UCMGs in the \INSPIRE\ sample. 

Nine out of the 52 systems stand out from the overall distribution, having DoR$>0.7$, which translates into very high fractions of mass formed within 3 Gyr from the BB ($>0.98$), and very short $t_{75}$ ($\le 1.5$) and $t_{\rm fin}$ ($\le3.5$).  
These are clearly the most extreme cases: they were completely formed 3 Gyr after the Big Bang. 
On the other extreme of the distribution, 7 UCMGs have formed 75\% of their stellar mass in a much longer time ($t_{75}>7$ Gyr), and therefore they cannot be considered relics of the ancient Universe. 
The remaining 36 systems span a large range of DoR ($0.25<$DoR$<0.7$). 

\begin{figure}
    \centering  \includegraphics[width=\columnwidth]{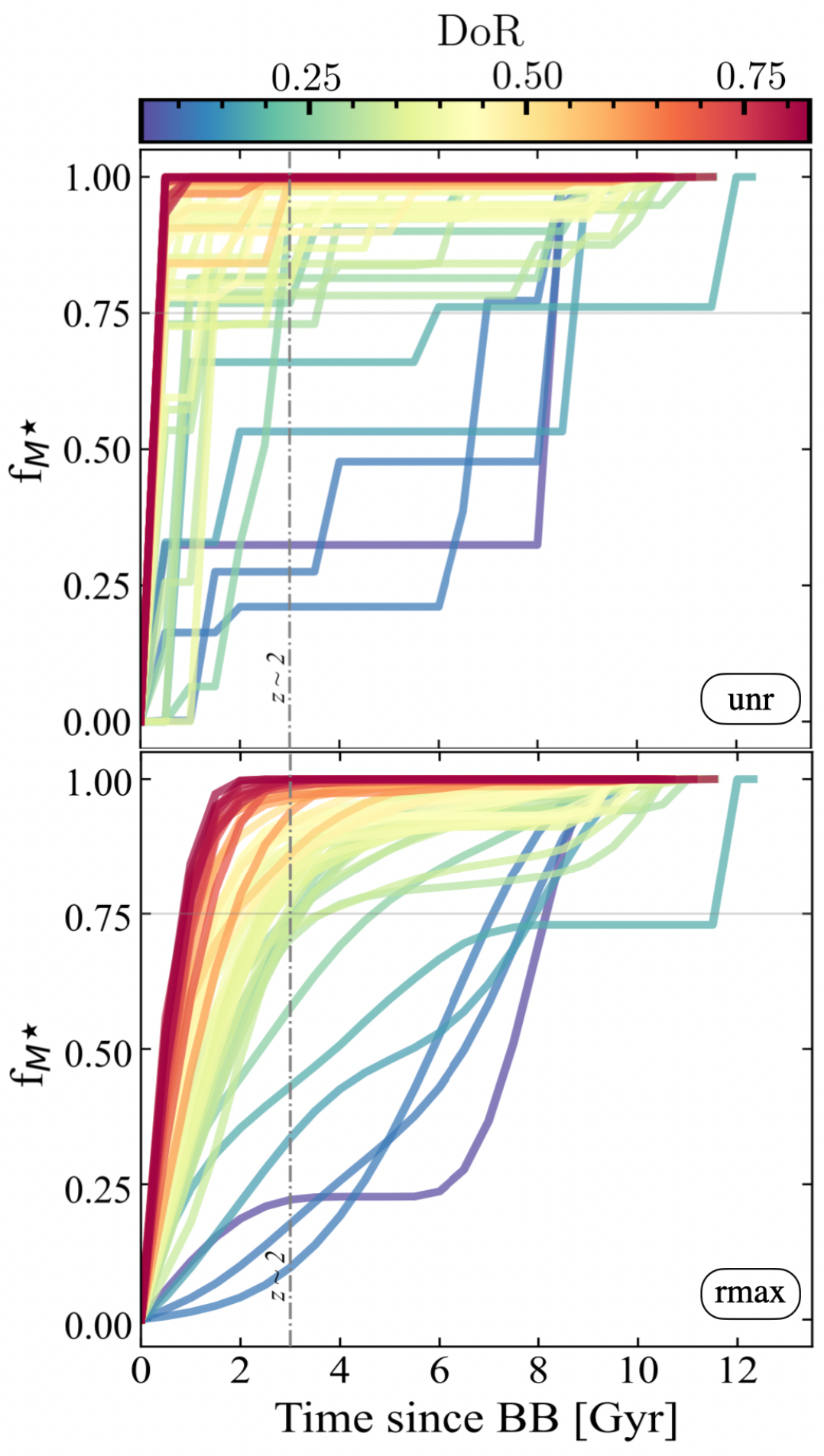}
    \caption{Cumulative fractions of mass formed in cosmic time since the BB for the unr (top) and the rmax (bottom) case for the entire \INSPIRE\ sample. The galaxies are colour-coded by DoR. }
    \label{fig:mass_fraction}
\end{figure}

Figure~\ref{fig:mass_fraction} shows the cumulative fraction of mass formed in cosmic time since the BB for the unregularised case (top) and the one with the maximum regularisation (bottom) for the entire \INSPIRE\ sample. The galaxies are colour-coded by their DoR, as indicated in the coulored bar on the top.  The figure highlights how different the SFHs of the 52 UCMGs are. All systems with DoR$\ge0.7$ (red and orange lines), have completely assembled their mass a couple of Gyr from the BB. These can be considered extreme relics.
Moving towards lower and lower DoR, the distribution of \Mfrac\ becomes instead more heterogeneous, in both the top and bottom panels. 
Interestingly, a similar diversity in the SFHs has also very recently been extended to compact galaxies at lower masses. For instance, \citet{Grebol2023} recently computed SFHs and formation timescales of compact galaxies with $9.9\le \log (M_{\star}/M_{\odot})\le 10$ from the 
Mapping Nearby Galaxies at APO (MaNGA) Survey \citep{Bundy15}. Using a machine learning grouping algorithm, they found three clearly distinct groups: galaxies that have remained mostly unchanged since their early formation epoch, galaxies with a more extended and continuous star formation and galaxies with an initial very low star-forming rate that than increases with time. 

%A clear difference in the way massive relic galaxies build their mass is what originally prompted the idea of a DoR from the three local ones \citep{Ferre-Mateu+17}. Moreover, this has been shown to also extend to compact galaxies at lower masses. In the recent paper of \citet{Grebol2023}, for instance, the SFHs and formation timescales of compact galaxies in {\tt MaNGA} were characterized. Then, following a machine learning approach, compact galaxies were classified in clearly distinctive groups.

%\chiara{Interestingly, four systems that have formed more than 90\% of their stellar mass during the first phase, have relatively lower DoR ($0.25<$DoR$<0.5$). These are objects that undergone a very late episode of star formation, which however only added few percent of the stellar mass, which is otherwise dominated by stars formed at $z>2$. Two other systems, instead, have a relatively low percentage (but still above 50\%) of mass formed during the first phase but slightly higher DoR than systems with similar M$_{\star,t_{\text{BB}}=3}$. The remaining 37 objects lie on a thigh line of increasing DoR with increasing M$_{\star,t_{\text{BB}}=3}$, as expected. }

%The vertical solid line in the plot, indicates instead a DoR$=0.5$. The 38 galaxies above this line have formed 75\% or more of their stellar mass by $z=2$.  They all have $t_{75} \le 3$ but could have much larger final assembly times. According to the relic operative definition used in \citetalias{Spiniello+21}, these objects are all classified as relics.  However, some of them have formed a minority of their fraction of stellar mass in very recent SF episodes. 

\begin{figure*}
    \centering  \includegraphics[width=\textwidth]{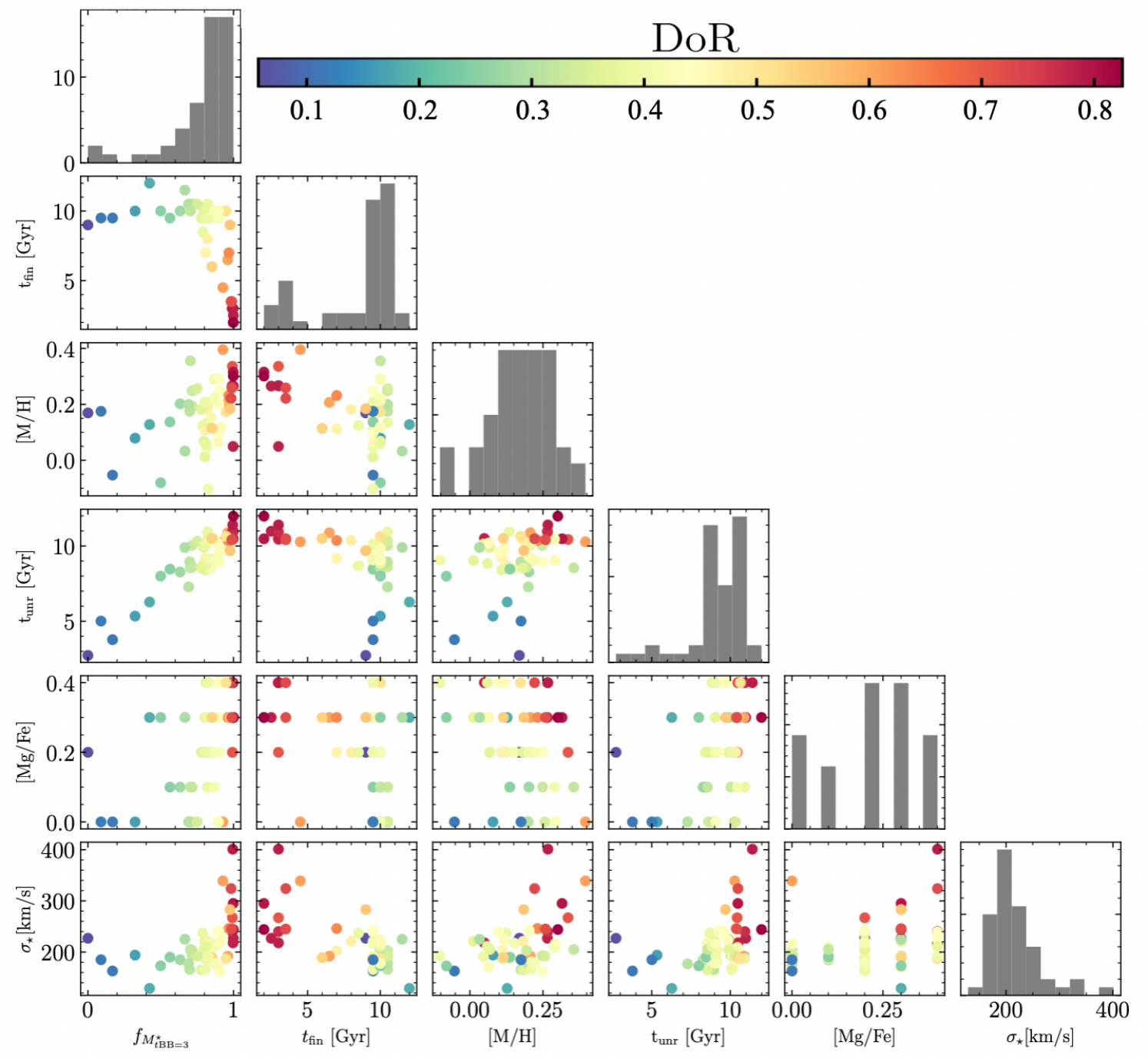}
    \caption{Correlation between the stellar population parameters. The \INSPIRE\ galaxies are colour-coded by the DoR, with red being the most extreme relics and blue indicating objects with a relatively more extended SFH. Galaxies with higher DoR are older (by definition), more metal-rich, and have higher [Mg/Fe] ratios and velocity dispersion values.}    
    \label{fig:corner}
\end{figure*}

\subsection{Relic characterisation}
The corner plot of Figure~\ref{fig:corner} shows the results of the stellar population analysis once all galaxies have been colour-coded by their DoR, as defined above. 
Extreme relics, with DoR$>0.7$ are plotted in red, while galaxies with a more extended SFH (i.e., non-relics, DoR$<0.25$) are plotted in blue. This is in line with the colour-coding used in previous \INSPIRE\ publications, but much more flexible.  
By construction, a higher DoR implies a lower $t_{\text{fin}}$ and a higher 
\Mfrac, i.e. an earlier final assembly with a higher fraction of stellar mass formed at $z>2$. 
Clearly, this also means that galaxies with higher DoR have older integrated ages, as visible from the plot. 

The 52 \INSPIRE\ UCMGs span a large range in stellar velocity dispersion ($\sim130-400$ \kms). For normal-sized galaxies, it is well known that the stellar population parameters and the star formation histories correlate with it, in the sense that galaxies with higher $\sigma_{\star}$ have overall older integrated ages and larger metallicities \citep[e.g.,][]{McDermid15}. This is true also for UCMGs\footnote{Comparing relics and their properties with ATLAS$^{\rm 3D}$ normal-size galaxies of similar masses (e.g. Fig.~22 in \citealt{Cappellari16}), the  \INSPIRE\ objects seems to have exactly the stellar population characteristics that one would expect by extrapolating the parameters of normal-size ETGs beyond the ZoE. At equal stellar masses, smaller galaxies are older, have larger stellar velocity dispersion values, richer metallicities and higher \afe.  Relics are the extreme tails of these distributions.}. However, we note that the definition of relics based on the cumulative mass fraction assembled in cosmic time does not directly depend on the stellar population parameters: UCMGs with very similar velocity dispersion (and also similar integrated age and metallicity)  can have very different DoR (e.g., J1228-0153 with DoR=0.39, $\sigma_{\star}=191\pm10$ and J1438-0127 with DoR=0.78, $\sigma_{\star}=218\pm11$).

%A relatively low DoR indicates instead that a large percentage of stars have been accreted or formed later on, hence the galaxies with low DoR are characterised by a much more time-extended star formation history. 

%We note that potentially,  if a galaxy has formed 100\% of its stellar mass in less than 1 Gyrs, the DoR>1. Given the many assumptions of the SSP method, that translate into uncertainties in the derived SFH, we have imposed a maximum of 1 to the DoR.  
An interesting result, which confirms what was already hinted at in \citetalias{Spiniello+21} 
is that 
%, is that a higher DoR also results in larger [M/H] and [Mg/Fe] abundances. Relics, and especially the most extreme ones, are characterised by metal-rich and $\alpha$-enhanced stellar populations. In addition, 
while galaxies with low DoR values are uniformly distributed around $\sigma\sim200-240$ \kms, there is a high fraction of UCMGs with high DoR that have larger $\sigma_{\star}$. This means that the stellar velocity dispersion can potentially be used, in the future, to select the most reliable relic candidates among UCMGs (see e.g., \citealt{Saulder+15_compacts}).  

\begin{figure}
    \centering  \includegraphics[width=\columnwidth]{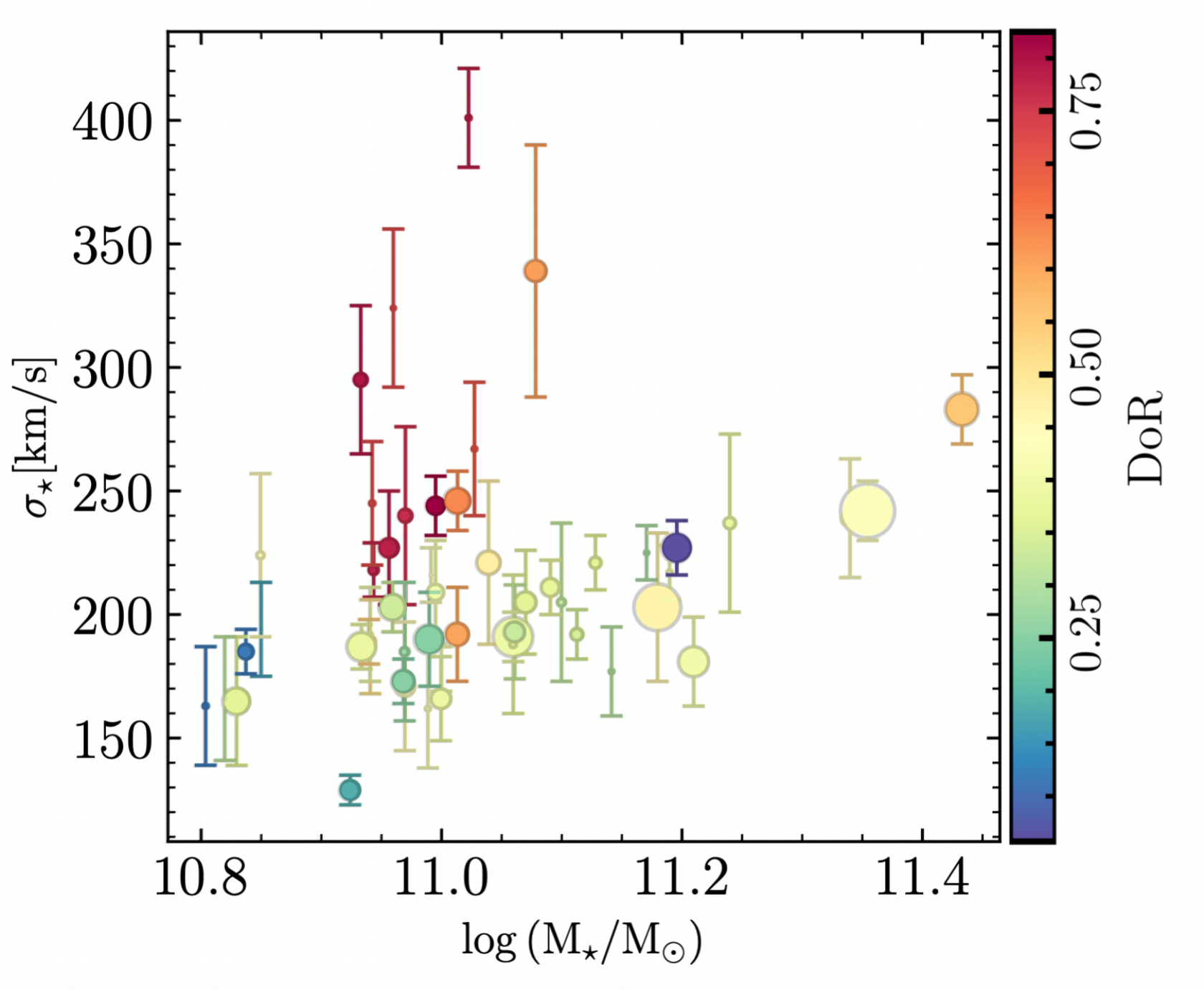}
    \caption{The M$_{\star}$--$\sigma$ relation for the \INSPIRE\ sample. Galaxies are colour-coded by their DoR and the symbol size scales with the \Reff. Objects with a high DoR (red) have overall larger stellar velocity dispersion compared to galaxies with lower DoR but equal stellar mass. }    
    \label{fig:mass-vdisp}
\end{figure}

Figure~\ref{fig:mass-vdisp} shows the stellar mass versus the stellar velocity dispersion. The galaxies are colour-coded by their DoR, as in the previous figure. 
We have already shown in Figure~12 of \citetalias{Spiniello+21} and Figure~10 of \citetalias{DAgo23} that at equal stellar mass, relics tend to have, on average, larger stellar velocity dispersion values than non-relics and than normal-sized galaxies. 
Here we do not wish to compare normal-sized galaxies with UMCGs, but only galaxies of very similar sizes but different DoR. Thanks to the larger number statistics, we now see that although some of the less extreme relics (i.e., yellow and orange points, with DoR$\sim 0.3-0.6$) have similar $\sigma_{\star}$ than non-relics of similar stellar mass and size, the majority of the systems with DoR$>0.7$ (red points) have larger velocity dispersions. 
Being the $\sigma_{\star}$ a proxy for the total mass of a system, given the virial theorem, and since all the plotted galaxies have very similar sizes and stellar masses, this plot implies that objects with a high DoR, have larger mass-to-light ratios than objects with lower DoR. 
One of the possible physical motivations for this could be the fact that these objects are characterised by a steeper low-mass end of the IMF slope, since dwarf stars contribute substantially to the mass but only add a few percent to the optical light \citep{Conroy13}. This seems to be the case, as we showed in \citet{Martin-Navarro+23}, although with very limited number statistics (5 relics and 5 non-relics). Alternatively, relics could have more dark matter or dark haloes with a higher central mass density. However, this seems to be in strong contrast with recent results showing that NGC1277, one of the three very well-studied extreme relics in the local Universe, is dark matter deficient up to 5 \Reff\ \citep{Comeron23}. 
We will tackle this problem in future dedicated publications i) precisely constraining the IMF and ii) performing a detailed dynamical model of the galaxies in the \INSPIRE\ final sample. 
We finally note that it is not surprising that we do not find extreme relics at very high masses, as it is very rare for such massive galaxies to form in such short timescales.

\begin{table*}
\caption{Results of the stellar population analysis. The first block of columns after the ID lists the ``SSP-equivalent" [Mg/Fe] abundances, and the mass-weighted ages and metallicities in the unregularised case (unr) and using the maximum regularisation allowed by the data (rmax). Uncertainties are obtained via bootstrapping. 
In the third block of columns, we list the minimum fraction of stellar mass assembled by $z=2$, the cosmic time at which the galaxies had formed 75\% of their mass, the cosmic time of final assembly and the age of the Universe at the redshift of the galaxies. Galaxies are ordered from the highest (most extreme relic) to the lowest (extended SFH) DoR, which are listed in the last column and the horizontal line separates the extreme relics. }
\label{tab:stelpop}
\begin{tabular}{l|ccccccc|cccc|c}
\hline
  \multicolumn{1}{c|}{ID} &
  \multicolumn{1}{c}{[Mg/Fe]} &
  \multicolumn{1}{c}{$t_{\mathrm{unr}}$} &
  \multicolumn{1}{c}{$t_{\mathrm{rmax}}$} &
  \multicolumn{1}{c}{$\Delta t$} &
  \multicolumn{1}{c}{[M/H]$_{\mathrm{unr}}$} &
  \multicolumn{1}{c}{[M/H]$_{\mathrm{rmax}}$} &
  \multicolumn{1}{c|}{$\Delta$[M/H]} &
 \multicolumn{1}{c}{\Mfrac} &   
 \multicolumn{1}{c}{$t_{75}$ } &
  \multicolumn{1}{c}{$t_{\text{fin}}$ } & 
  \multicolumn{1}{c}{$t_{\mathrm{Uni, z}}$} &
  \multicolumn{1}{|c}{DoR} \\
  \multicolumn{1}{c|}{KiDS} &
  \multicolumn{1}{c}{ ($\pm0.1$ dex) } &
  \multicolumn{1}{c}{(Gyr)} &
  \multicolumn{1}{c}{(Gyr)} &
  \multicolumn{1}{c}{(Gyr)} &
  \multicolumn{1}{c}{dex} &
  \multicolumn{1}{c}{dex} &
  \multicolumn{1}{c|}{dex} &
  \multicolumn{1}{c}{ } &
  \multicolumn{1}{c}{(Gyr)} &
  \multicolumn{1}{c}{(Gyr)} &  
  \multicolumn{1}{c}{(Gyr)} &
  \multicolumn{1}{|c}{} \\
\hline\hline
  J0847+0112 & 0.3 & 12.0 & 11.6 & 0.2 & 0.30  & 0.35  & 0.02     & $0.9988\pm0.0002$ & $0.9 \pm0.3$ & $2.0\pm0.7$ & 11.52 & 0.83\\  %$1.00$ ,
  J2305-3436 & 0.3$^*$ & 10.5 & 10.1 & 0.2 & 0.32  & 0.34  & 0.02 & $0.9980\pm0.0007$ & $1.0 \pm0.3$ & $2.0\pm0.7$ & 10.29 & 0.80\\  %$0.94$ ,
  J0909+0147 & 0.4 & 11.4 & 11.1 & 0.3 & 0.27  & 0.31  & 0.02     & $0.9950\pm0.0013$ & $0.9 \pm0.2$ & $3.0\pm4.0$ & 11.11 & 0.79\\  %$0.90$ ,
  J2204-3112 & 0.3 & 11.0 & 10.6 & 0.3 & 0.27  & 0.30  & 0.02     & $0.9992\pm0.0003$ & $1.0 \pm0.4$ & $2.5\pm0.9$ & 10.67 & 0.78\\  %$0.88$ ,
  J1438-0127 & 0.4 & 10.5 & 10.1 & 0.4 & 0.05  & 0.11  & 0.05     & $0.9971\pm0.0013$ & $0.9 \pm0.3$ & $3.0\pm1.1$ & 10.40 & 0.78\\  %$0.88$ ,
  J1040+0056 & 0.4 & 10.9 & 10.6 & 0.3 & 0.27  & 0.28  & 0.02     & $0.991\pm0.003$   & $1.0 \pm0.3$ & $3.0\pm3.9$ & 10.54 & 0.77\\  %$0.85$ ,
  J0842+0059 & 0.4 & 10.5 & 10.0 & 0.2 & 0.22  & 0.26  & 0.02     & $0.983\pm0.007$   & $1.1 \pm0.3$ & $3.5\pm1.8$ & 10.31 & 0.73\\  %$0.77$ ,
  J0211-3155 & 0.3 & 10.4 & 9.9  & 0.2 & 0.26  & 0.28  & 0.03     & $0.989\pm0.003$   & $1.1 \pm0.3$ & $3.5\pm4.0$ & 10.26 & 0.72\\  %$0.75$ ,
  J2359-3320 & 0.2 & 10.4 & 9.7  & 0.2 & 0.34  & 0.33  & 0.03     & $0.993\pm0.002$   & $1.4 \pm0.4$ & $3.0\pm3.9$ & 10.38 & 0.71\\  %$0.70$ ,
\hline                     J0920+0212 & 0.3 & 10.4 & 9.9  & 0.3 & 0.23  & 0.26  & 0.03     & $0.968\pm0.012$   & $0.9 \pm0.2$   & $7.0\pm3.8$ & 10.46 & 0.64\\  %$0.75$ ,
  J1412-0020 & 0.0 & 10.3 & 9.3  & 0.3 & 0.40  & 0.32  & 0.03     & $0.93\pm0.03$     & $1.9 \pm0.6$ & $4.5\pm4.1$ & 10.48 & 0.61\\  %$0.54$ ,
  J1449-0138 & 0.3 & 10.9 & 10.2 & 0.3 & 0.21  & 0.25  & 0.03     & $0.961\pm0.016$   & $1.3 \pm0.3$ & $6.5\pm3.8$ & 10.60 & 0.60\\  %$0.62$ ,
  J0224-3143 & 0.3 & 9.8  & 9.4  & 0.3 & 0.18  & 0.21  & 0.03     & $0.976\pm0.004$   & $0.9 \pm0.2$ & $9.0\pm3.7$ & 9.53  & 0.56\\  %$0.75$ ,
  J0838+0052 & 0.3$^*$ & 10.5 & 9.8  & 0.5 & 0.11  & 0.13  & 0.04 & $0.85\pm0.05$     & $1.9 \pm0.6$ & $6.0\pm3.3$ & 10.56 & 0.54\\  %$0.49$ ,
  J0317-2957 & 0.4 & 10.6 & 10.3 & 0.3 & 0.17  & 0.19  & 0.04     & $0.949\pm0.018$   & $1.1 \pm0.3$ & $10.0\pm4.0$ & 10.64 & 0.51\\  %$0.63$ ,
  J2202-3101 & 0.2 & 10.4 & 9.5  & 0.5 & 0.18  & 0.22  & 0.03     & $0.82\pm0.07$     & $1.5 \pm0.4$ & $8.0 \pm2.9$ & 10.10 & 0.48\\  %$0.53$ ,
  J1457-0140 & 0.2 & 9.2  & 8.5  & 0.5 & 0.11  & 0.12  & 0.05     & $0.81\pm0.07$     & $2.1 \pm0.7$ & $7.0 \pm2.7$ & 9.93  & 0.47\\  %$0.45$ ,
  J0844+0148 & 0.4 & 9.8  & 9.1  & 0.1 & 0.12  & 0.17  & 0.05     & $0.91\pm0.03$     & $1.5 \pm0.4$ & $10.0\pm4.0$ & 10.43 & 0.45\\  %$0.52$ ,
  J1218+0232 & 0.3 & 10.2 & 9.3  & 0.4 & 0.06  & 0.14  & 0.05     & $0.89\pm0.04$     & $1.5 \pm0.5$ & $9.5 \pm3.3$ & 10.20 & 0.45\\  %$0.52$ ,
  J2356-3332 & 0.2 & 8.9  & 7.9  & 0.7 & 0.25  & 0.29  & 0.05     & $0.90\pm0.02$     & $1.7 \pm0.6$ & $9.5 \pm3.9$ & 9.92  & 0.44\\  %$0.49$ ,
  J0917-0123 & 0.4 & 9.6  & 8.4  & 0.3 & 0.10  & 0.18  & 0.04     & $0.89\pm0.04$     & $1.6 \pm0.5$ & $9.5 \pm3.9$ & 9.73  & 0.44\\  %$0.51$ ,
  J0918+0122 & 0.2$^*$ & 9.3  & 8.6  & 0.3 & 0.20  & 0.23  & 0.05 & $0.90\pm0.03$     & $1.6 \pm0.5$ & $9.5 \pm4.0$ & 9.62  & 0.43\\  %$0.51$ ,
  J0240-3141 & 0.0 & 9.1  & 8.5  & 0.7 & 0.29  & 0.29  & 0.02     & $0.90\pm0.03$     & $1.9 \pm0.6$ & $10.0\pm3.9$ & 10.47 & 0.43\\  %$0.47$ ,
  J0314-3215 & 0.2$^*$ & 9.0  & 8.1  & 0.5 & 0.22  & 0.26  & 0.05 & $0.87\pm0.02$     & $1.8 \pm0.6$ & $10.0\pm4.1$ & 10.39 & 0.42\\  %$0.47$ ,
  J1420-0035 & 0.1 & 10.5 & 9.2  & 0.5 & 0.29  & 0.30  & 0.03     & $0.87\pm0.05$     & $2.2 \pm0.8$ & $10.0\pm3.9$ & 10.77 & 0.41\\  %$0.43$ ,
  J1411+0233 & 0.2 & 8.5  & 7.8  & 0.5 & 0.15  & 0.18  & 0.04     & $0.84\pm0.04$     & $1.7 \pm0.5$ & $9.5 \pm4.1$ & 9.74  & 0.41\\  %$0.48$ ,
  J1114+0039 & 0.2 & 9.2  & 8.2  & 0.5 & 0.21  & 0.25  & 0.03     & $0.86\pm0.04$     & $1.9 \pm0.6$ & $10.0\pm3.6$ & 10.27 & 0.40\\  %$0.45$ ,
  J0316-2953 & 0.4$^*$ & 9.1  & 8.0  & 0.3 & -0.10 & 0.02  & 0.08 & $0.83\pm0.06$     & $1.9 \pm0.5$ & $9.5 \pm4.1$ & 9.74  & 0.40\\  %$0.45$ ,
  J1228-0153 & 0.3 & 10.1 & 9.0  & 0.6 & 0.05  & 0.09  & 0.05     & $0.82\pm0.08$     & $2.3 \pm0.7$ & $9.5 \pm3.8$ & 10.3  & 0.39\\  %$0.41$ ,
  J0857-0108 & 0.2 & 10.1 & 9.1  & 0.4 & 0.21  & 0.21  & 0.04     & $0.80\pm0.05$     & $2.6 \pm0.9$ & $9.5 \pm4.0$ & 10.56 & 0.39\\  %$0.39$ ,
  J1447-0149 & 0.1 & 10.9 & 9.8  & 0.7 & 0.13  & 0.18  & 0.04     & $0.82\pm0.07$     & $2.5 \pm0.9$ & $10.5\pm4.3$ & 11.19 & 0.38\\  %$0.39$ ,
  J1527-0012 & 0.2 & 8.7  & 7.7  & 0.6 & 0.124 & 0.18  & 0.05     & $0.79\pm0.05$     & $2.7 \pm0.9$ & $8.5 \pm3.2$ & 9.4   & 0.38\\  %$0.39$ ,
  J1527-0023 & 0.2 & 8.4  & 7.4  & 0.9 & 0.11  & 0.12  & 0.09     & $0.81\pm0.06$     & $2.4 \pm0.8$ & $9.5 \pm3.4$ & 9.82  & 0.37\\  %$0.40$ ,
  J0321-3213 & 0.4 & 8.8  & 8.7  & 0.4 & 0.06  & 0.04  & 0.05     & $0.81\pm0.06$     & $2.4 \pm0.7$ & $10.0\pm3.7$ & 10.32 & 0.37\\  %$0.40$ ,
  J2312-3438 & 0.3 & 9.0  & 8.6  & 0.4 & 0.012 & 0.03  & 0.04     & $0.80\pm0.07$     & $2.5 \pm0.7$ & $9.5 \pm3.9$ & 9.68  & 0.36\\  %$0.39$ ,
  J1414+0004 & 0.2 & 9.9  & 8.4  & 0.5 & 0.07  & 0.14  & 0.06     & $0.80\pm0.08$     & $2.5 \pm0.9$ & $10.0\pm4.1$ & 10.24 & 0.36\\  %$0.39$ ,
  J1202+0251 & 0.2 & 8.5  & 7.3  & 0.7 & 0.18  & 0.18  & 0.06     & $0.78\pm0.06$     & $2.6 \pm0.9$ & $9.5 \pm3.1$ & 10.0  & 0.36\\  %$0.38$ ,
  J1128-0153 & 0.0 & 10.3 & 9.4  & 0.5 & 0.26  & 0.25  & 0.02     & $0.75\pm0.09$     & $2.9 \pm1.1$ & $10.5\pm4.3$ & 11.04 & 0.34\\  %$0.35$ ,
  J1436+0007 & 0.1 & 10.0 & 8.4  & 0.8 & 0.25  & 0.27  & 0.03     & $0.72\pm0.09$     & $3.1 \pm1.1$ & $10.5\pm4.4$ & 11.05 & 0.33\\  %$0.33$ ,
  J1417+0106 & 0.0 & 8.6  & 8.0  & 0.6 & 0.19  & 0.18  & 0.03     & $0.72\pm0.08$     & $3.6 \pm1.2$ & $10.5\pm3.6$ & 11.49 & 0.33\\  %$0.31$ ,
  J0904-0018 & 0.1 & 8.6  & 7.8  & 0.5 & 0.36  & 0.28  & 0.04     & $0.70\pm0.10$     & $3.2 \pm1.1$ & $10.0\pm3.8$ & 10.28 & 0.32\\  %$0.33$ ,
  J1402+0117 & 0.0 & 9.0  & 8.5  & 0.9 & 0.17  & 0.17  & 0.05     & $0.70\pm0.07$     & $3.2 \pm1.1$ & $10.5\pm4.1$ & 10.72 & 0.31\\  %$0.33$ ,
  J1156-0023 & 0.0 & 7.3  & 6.7  & 0.8 & 0.20  & 0.20  & 0.03     & $0.69\pm0.06$     & $3.6 \pm1.2$ & $10.5\pm4.1$ & 10.7  & 0.30\\  %$0.31$ ,
  J1026+0033 & 0.3 & 9.9  & 9.3  & 0.4 & 0.03  & 0.03  & 0.03     & $0.67\pm0.05$     & $3.6 \pm1.5$ & $11.5\pm3.2$ & 11.55 & 0.29\\  %$0.30$ ,
  J2257-3306 & 0.1 & 8.3  & 8.5  & 0.7 & 0.20  & 0.21  & 0.03     & $0.63\pm0.12$     & $8.1 \pm3.1$ & $10.0\pm4.2$ & 10.68 & 0.27\\  %$0.24$ ,
  J0920+0126 & 0.1$^*$ & 8.5  & 7.0  & 0.9 & 0.14  & 0.14  & 0.05 & $0.56\pm0.13$     & $8.5 \pm3.4$ & $9.5 \pm3.9$ & 10.16 & 0.25\\  %$0.22$ ,
  J0326-3303 & 0.3 & 8.0  & 8.4  & 0.6 & -0.08 & -0.09 & 0.05     & $0.50\pm0.17$     & $3.3 \pm1.2$ & $10.0\pm3.5$ & 10.3  & 0.25\\  %$0.29$ ,
  J1142+0012 & 0.3 & 6.3  & 5.3  & 0.6 & 0.13  & 0.11  & 0.04     & $0.42\pm0.12$     & $11.5\pm4.8$ & $12.0\pm3.3$ & 12.32 & 0.18\\  %$0.17$ ,
  J1456+0020 & 0.0 & 5.3  & 5.3  & 1.1 & 0.08  & 0.12  & 0.06     & $0.32\pm0.09$     & $8.7 \pm3.4$ & $10.0\pm2.1$ & 10.52 & 0.17\\  %$0.17$ ,
  J1154-0016 & 0.0 & 3.8  & 3.8  & 0.8 & -0.05 & -0.05 & 0.07     & $0.17\pm0.14$     & $8.3 \pm3.4$ & $9.5 \pm3.3$ & 9.95  & 0.11\\  %$0.15$ ,
  J0226-3158 & 0.0 & 5.0  & 5.3  & 1.0 & 0.18  & 0.21  & 0.04     & $0.09\pm0.30$     & $7.1 \pm3.2$ & $9.5 \pm2.5$ & 10.9  & 0.12\\  %$0.14$ ,
  J2327-3312 & 0.2 & 2.7  & 2.7  & 0.5 & 0.17  & 0.16  & 0.04     & $0.00\pm0.18$     & $8.4 \pm3.5$ & $9.0 \pm0.4$ & 9.34  & 0.06\\  %$0.12$ ,
\hline\hline
\end{tabular}                              
\begin{flushright}
$^*$For these systems the new estimate is $0.1\times$ higher/lower than that of DR1.
\end{flushright}

\end{table*}

\section{Summary and Conclusions}
\label{sec:conclusions}
In this paper, we have presented the complete sample of the Investigating Stellar Populations In RElics (\INSPIRE) Survey. 
The goal of \INSPIRE\ has been to 
study the kinematics and stellar population parameters of a sample of UCMGs with effective radii \Reff$<2$ kpc and stellar masses \Mstar$>6\times10^{10}$ \Msun to  
build the first sizeable catalogue of spectroscopically confirmed relics at $0.1<z<0.4$. 
Observations were based on an ESO Large Program, which was completed in March 2023. The data have been made available to the community through three ESO Data Releases (DR1: \citealt{Spiniello+21}, DR2: \citealt{DAgo23}), the last of which has been described in this paper. 

For each of the 52 UCMGs we have derived highly accurate spectroscopic redshifts, precise estimates of the integrated stellar velocity dispersion and the associated uncertainties, ``SSP-equivalent" [Mg/Fe] from line-indices and mass-weighted stellar ages and metallicities from full-spectral fitting.  All these quantities are released in a publicly available catalogue, together with morpho-photometric characteristics derived from KiDS and VIKING multi-band images. 

In this paper, we have: 
\begin{itemize}
    \item[-] added 12 new, still unpublished systems to the previous DRs, releasing their UVB, VIS and NIR spectra, as well as the combined spectrum smoothed at a resolution of FWHM$=2.51$\AA, that of the SSP models. All spectra have been extracted from an aperture that encapsulates 50\% of the light, but that takes into account that the data are seeing dominated, 
    \item[-] (re-)computed the stellar velocity dispersion ($\sigma_{\star}$) for all the \INSPIRE\ galaxies using the code \ppxf\ with a uniform configuration, %(ADEGREE=13, CLEAN=TRUE, $\lambda=[3000-10000]$\AA). 
    \item[-] obtained stellar ages, metallicities and [Mg/Fe] abundances for 33 new systems (DR2 and DR3) and for the 19 systems already presented in DR1 from the smoothed 1D UVB+VIS spectra. %at $\lambda=[3525-7500]$ \AA, %This wavelength range allows us to keep fix the IMF slope.     
\end{itemize}

These quantities allowed us to derive the SFHs of the galaxies in the entire \INSPIRE\ sample and hence define, for each of the UCMGs, a degree of relicness (DoR), based on the fraction of stellar mass assembled 3 Gyr after the BB ($z\ge2$), on the time at which the galaxy had formed 75\% of its stellar mass and on the cosmic time of final assembly (100\% of the stellar mass formed), taking into account the redshifts of the systems. 
The DoR is defined as a dimensionless value that varies from 1 (the most extreme case: a galaxy that has fully assembled 700 Myrs after the BB) to 0 (a galaxy that is still forming stars today). 

Among the 52 \INSPIRE\ UCMGs, 9 can be defined as ``extreme relics", since they had formed >99\% of their stellar mass by $z = 2$, corresponding to a DoR$ > 0.7$. 
According to the simpler definition given in \citetalias{Spiniello+21}, 38 objects would be classified as relics, having formed more than 75\% of their stellar mass by $z=2$ ($0.34\le$DoR$\le0.64$). 
This number nicely confirms (and even increases) the initial expectation of the survey which predicted that 50\% of the sample would be classified as relic, following this simple, operational definition. 
However, here, in this final data release, we have defined a much more informative classification of these UCMGs, assigning a DoR to each system, that takes into account these SFH parameters in a roubst way. 
The DoR is found to correlate with the stellar velocity dispersion, [Mg/Fe], metallicity,  and (by construction) age. 

Among the 38 relics, the different DoR
represents different SFHs, with some systems that have formed all their stars just 2 Gyrs
after the Big Bang (extreme, $DoR>0.7$) and some others with much longer $t_{\rm fin}$ ($0.34\le$DoR$\le0.64$).
The remaining 14 UCMGs are instead systems that are still forming stars or that stopped much later on in cosmic time. These objects could pose problems to simulations that will have to
explain the presence of multiple star formation episodes but a lack of growth. We note however that these
systems, or at least a fraction of them, could also be “stripped” objects that were much bigger in size
but then lost part of their materials. Unfortunately with ground-based observations and \INSPIRE\ integrated and seeing-dominated spectra we cannot disentangle between these cases. 

%Finally, we have also inferred the lower limit to the number density of all UCMGs and relics from the entire \INSPIRE\ sample. We find $\log \rho \sim 3.9 \times 10^{-7} \text{Mpc}^{-3}$ for the 52 objects, and $\log \rho \sim 2.8 \times 10^{-7} \text{Mpc}^{-3}$ when restricting the computation to the 38 systems with \Mfrac$\ge 0.75$. These numbers are slightly smaller than the value reported by \citetalias{Ferre-Mateu+17} and than the predictions from simulations. However, since we can not take into account the completeness and the selection function of the UCMGs/relics, the inferred numbers have to be considered as lower limits only. 

In conclusion, the \INSPIRE\ Survey has successfully built a sizeable catalogue of relics and younger UCMGs, which is released to the community. This opens up the opportunity to study in great detail the first phases of formation and mass assembly of massive galaxies at high-$z$, but with the ease of being in the nearby Universe.

%The catalogue, as well as all the \INSPIRE\ spectra, are publicly available through the ESO Science Portal, as described in the Data Avalaibility disclaimer below. 

%%%%%%%%%%%%%%%%%%%%%%%%%%%%%%%%%%%%%%%%%%%%%%%%%%
\section*{Data Availability}
The data described in this paper are publicly available via the ESO Phase 3 
Archive Science Portal under the collection \INSPIRE (\url{https://archive.eso.org/scienceportal/home?data_collection=INSPIRE}).

\section*{Acknowledgements}
%CS is supported by an `Hintze Fellow' at the Oxford Centre for Astrophysical Surveys, which is funded through generous support from the Hintze Family Charitable  Foundation.  GD acknowledges support by ANID, BASAL, FB210003. 
CS, CT, FLB, DB, PS, AG, and SZ  acknowledge funding from the INAF PRIN-INAF 2020 program 1.05.01.85.11. AFM acknowledges support from RYC2021-031099-I and PID2021-123313NA-I00 of MICIN/AEI/10.13039/501100011033/
FEDER,UE,NextGenerationEU/PRT. DS is supported by JPL, which is operated under a contract by Caltech for NASA. 
%DS is a member of the International Max Planck Research School (IMPRS) for Astronomy and Astrophysics at the Universities of Bonn and Cologne.
The authors wish to thank the ESO Archive Science Group for the great support with the Data Release.



\bibliographystyle{mnras}
\bibliography{INSPIRE} 

%%%%%%%%%%%%%%%%%%%%%%%%%%%%%%%%%%%%%%%%%%%%%%%%%%

%%%%%%%%%%%%%%%%% APPENDICES %%%%%%%%%%%%%%%%%%%%%

\appendix
\section{Comparison with DR1}
\label{app:comparison}
In this Appendix, we show a direct comparison between the stellar population results obtained in DR1 \citep{Spiniello+21} and these presented here for 19 objects. 
For consistency, we only consider here the optimally extracted spectra from DR1, the only one we use in this DR3.

Here below we briefly list the main (small) differences between DR1 and this paper. 
\begin{itemize}
    \item \textit{Fitting region:} in DR1, we restrict the fit to the spectral region [3500-7000]\AA, while here we go slightly redder, up to 7500\AA. The spectral range used for the fit does play a role both in measuring the velocity dispersion (see the tests in \citetalias{DAgo23}) and in  determining the SFHs for a couple of systems with low DoR. This is probably caused by the presence of a noisy region around 7100-7200\AA\ for them (this is also visible from the bottom panel of Figure~\ref{fig:ppxf_res}, showing J2327-3312, the youngest galaxy in the \INSPIRE\ sample.  
    However, this does not affect the main conclusions nor the classification obtained in the paper. 
    \item \textit{\ppxf\ version:} the version 7.4 was used in DR1, whereas version 8.2.4 is used in this paper. 
    \item \textit{Cleaning and masking:} we slightly improve the masking of bad pixel regions and additionally used the CLEAN keyword in \ppxf.   
\end{itemize}

\begin{figure}
    \centering  \includegraphics[width=\columnwidth]{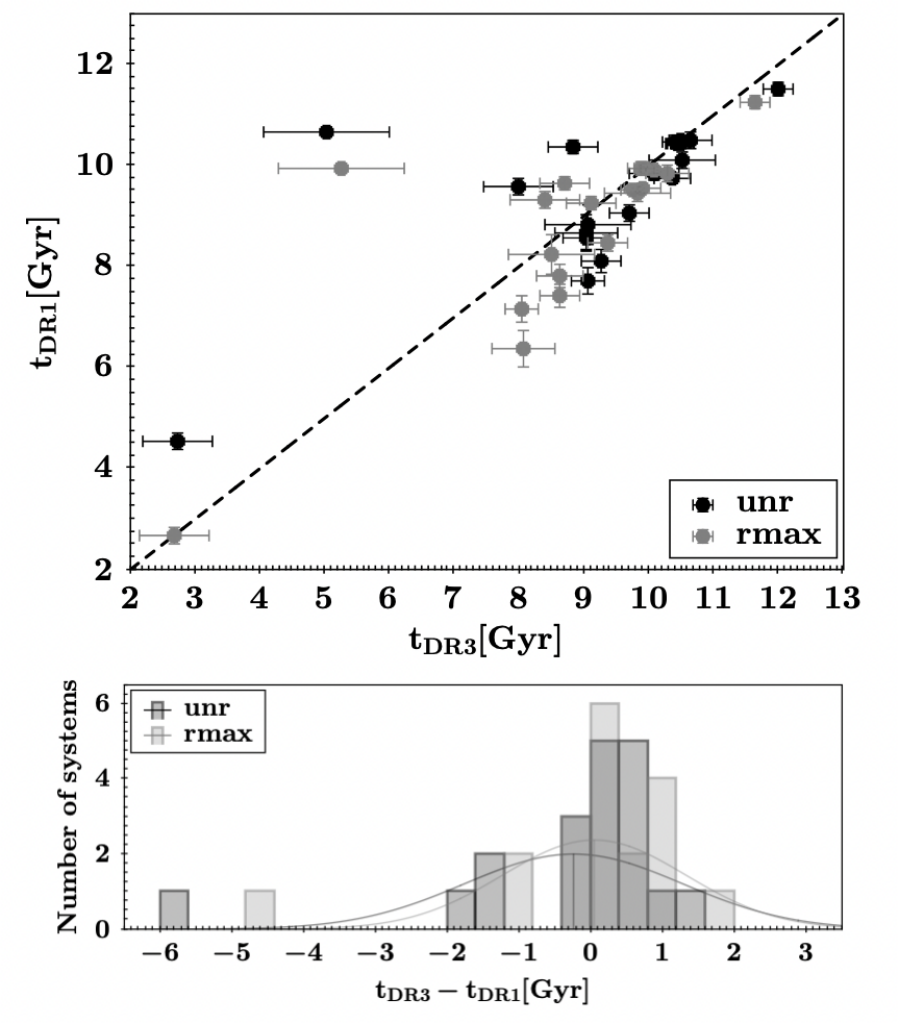}
    \caption{Comparison between DR3 and DR1 stellar ages. The top panel shows the scatter plot, with the 1-to-1 relation shown as a dotted line.  The bottom panel shows the histogram of the difference between the two values. Different colours indicate the unr and the rmax estimates. }
    \label{fig:comparison_age}
\end{figure}

\begin{figure}
    \centering  \includegraphics[width=\columnwidth]{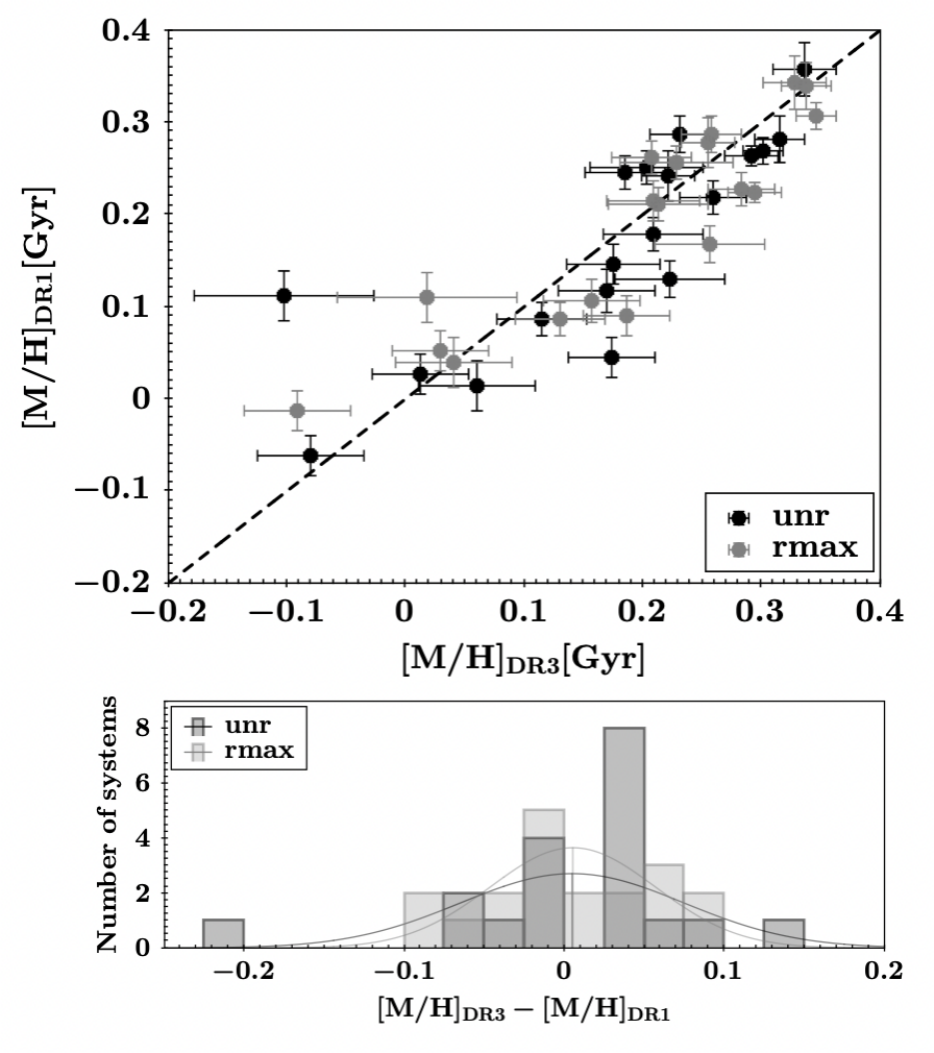}
    \caption{Same as in Figure~\ref{fig:comparison_age} but with the metallicity estimates. }
    \label{fig:comparis_met}
\end{figure}

Overall, a fair agreement is found for the majority of the systems for both age and metallicity,  with however a small systematic effect towards slightly older ages,  as shown in the top panels of Figure~\ref{fig:comparison_age} and Figure~\ref{fig:comparis_met}.  
The bottom panels of each plot the histograms of the difference between the age and metallicity computed in the two DRs.  For both the unr and the rmax cases,  a Gaussian fit to the difference shows no systematic differences and a spread of about 1-2 Gyr for the age and of $\sim-0.1$ for [M/H].  
With the exception of two galaxies: 

\begin{itemize}
    \item {\bf J0226-3158}. For this galaxy, a much older stellar age was inferred ($\sim 10$ Gyr) in DR1, than the one we infer here ($\sim6$ Gyr), for both the unregularised and the regularised cases. 

    \item {\bf J0316-2953}. For this galaxy,  the age estimates agree within $1.5 \sigma$ (with the ones estimated here being $\sim1$ Gyr older), but the inferred [M/H] values differ by 0.2 dex.  In DR1 J0316-2953 resulted to be slightly super-solar,  while here it has [M/H] values below solar now with the new analysis. The disagreement is however smaller when considering the rmax case. 
\end{itemize}

\section{UCMGs and relics number density}
\label{sec:results_numbdens}

With the final \INSPIRE\ sample at our disposal and the relic confirmation described above, we can update the lower limit to the number density of relics (and UCMGs) obtained in previous DRs. This number provides an important constraint for galaxy assembly models and simulations. 

Following the approach of \citetalias{Tortora+18_UCMGs}, already used also in \citetalias{Spiniello+21}, the number density is defined as
\begin{equation}
\rho = \dfrac{f_{\text{area}} \times N_{\text{conf}}}{V_{\text{KiDS,z}}}
\end{equation}
where $N_{\text{conf}}$ is the number of objects. We considered both the 38 systems that have \Mfrac$\ge0.75$ and the entire sample of 52 UCMGs. 
The $V_{\text{KiDS,z}}$ is the cosmic comoving volume contained within the redshift window covered by \INSPIRE\ ($0.1<z<0.41$), and is calculated as: $V_{\text{KiDS,z}} = (V_{\text{z=0.41}}-V_{\text{z=0.1}})$. Finally,  $f_{\text{area}}$ is a normalization factor used to normalise the number of objects to the full sky area. It is defined as the ratio $A_{\text{sky}}/A_{\text{KiDS}}$, where $A_{\text{sky}} = 41253$ deg$^{2}$ is the full sky area and $A_{\text{KiDS}}=333$ deg$^2$ is the KiDS DR3 effective area within which the UCMGs are found \citep{deJong+17_KiDS_DR3}.

\begin{figure}
    \centering    \includegraphics[width=\columnwidth]{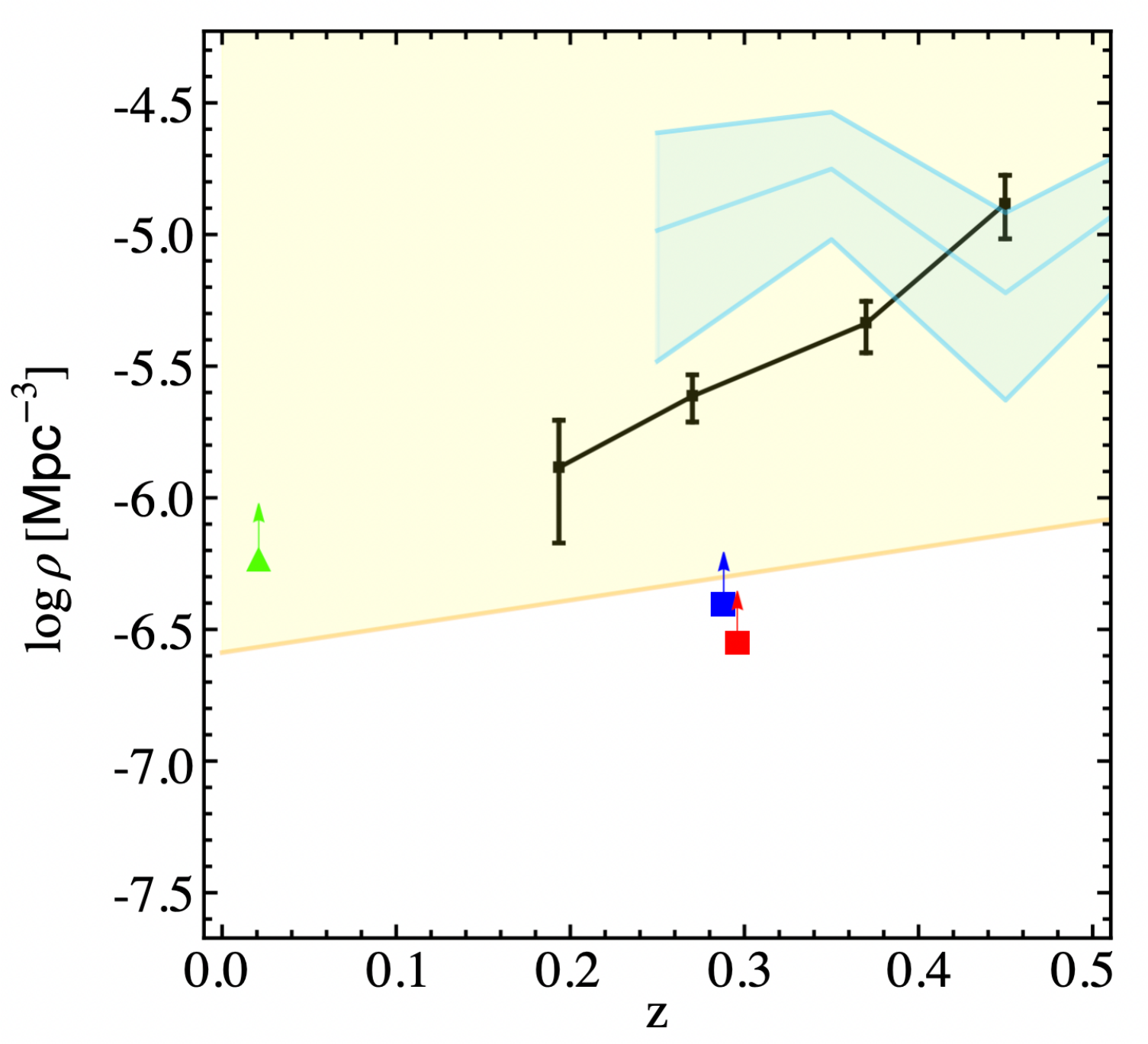}
    \caption{Number density of \INSPIRE\ UCMGs (blue square) and relics (\Mfrac$\ge 0.75$, red square) versus redshift. 
    The black points and lines are the UMCGs from \citetalias{Tortora+18_UCMGs}. The green  triangle shows the lower limit inferred from the three local relics by \citet{Ferre-Mateu+17}. The shaded yellow region highlights the region predicted by  simulations. This has been computed as the lowest limit presented in \citep{Quilis_Trujillo13}. 
    The cyan line and region show the number counts for compact galaxies in the COSMOS area \citep{Damjanov+15_compacts}. }
    \label{fig:numb_dens}
\end{figure}

The number density for the 38 relics (red point, $\log \rho \sim 2.8 \times 10^{-7} \text{Mpc}^{-3}$) and the entire \INSPIRE\ sample (blue point, $\log \rho \sim 3.9 \times 10^{-7} \text{Mpc}^{-3}$) is shown in Figure~\ref{fig:numb_dens}. This second value, while does not represent a benchmark for simulations wishing to compare the number of predicted relics, is useful for comparing our findings to the many works computing the number density of UCMGs, without any distinctions between old and younger systems (e.g., \citealt{Tortora+18_UCMGs, Scognamiglio20}). 
In the same figure, we also show the lower limit estimated by \citet{Ferre-Mateu+17} from the three local relics (green triangle) and the trend inferred by \citetalias{Tortora+18_UCMGs} (black points) considering all UCMGs found in KiDS. The yellow region shows the zone predicted by \citet{Quilis_Trujillo13}, who
used the semi-analytical simulations from \citet{Guo+11_sims} and \citet{Guo+13_sims}  to define “relic compacts”  galaxies with mass changing less than 10\% from $z\sim2$. A mild evolution is found with redshift for all the mentioned studies. The cyan region and line represent the number densities from \citet{Damjanov+15_compacts} 
inferred for galaxies in the COSMOS survey. In this case, no redshift evolution is found. 

Our estimates are slightly below the value reported by \citetalias{Ferre-Mateu+17} and the limit predicted by simulations. However, we stress that this number has to be considered as a lower limit on the relic number density, as the completeness and the selection function of the UCMGs/relics have not been taken into account.

\section{1D Spectra}
\label{app:spectra}
\begin{figure*}
    \centering  \includegraphics[width=0.9\textwidth]{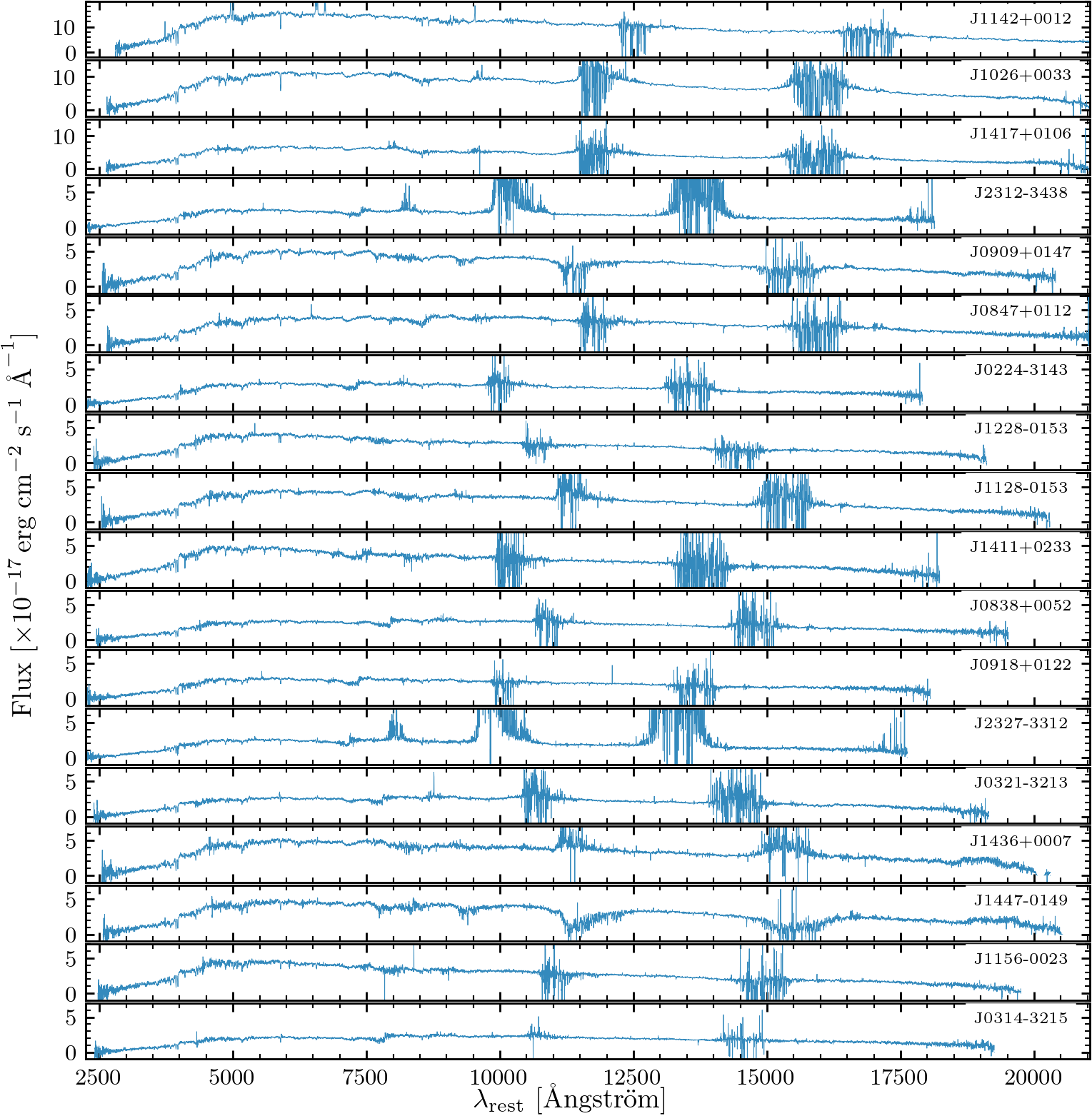}
    \caption{
    1D smoothed spectra (9-pixel wide median filter) for the systems in the high SNR group ($\mathrm{SNR}>35$). The single-arm spectra are joined together after restframing and rebinning to a final resolution of FWHM$=2.51$\AA. Residuals from telluric lines have not been removed from the spectra, but have been masked out for the kinematics and stellar population analysis.}
\label{fig:spectra_high}
\end{figure*}

\begin{figure*}
    \centering  \includegraphics[width=0.9\textwidth]{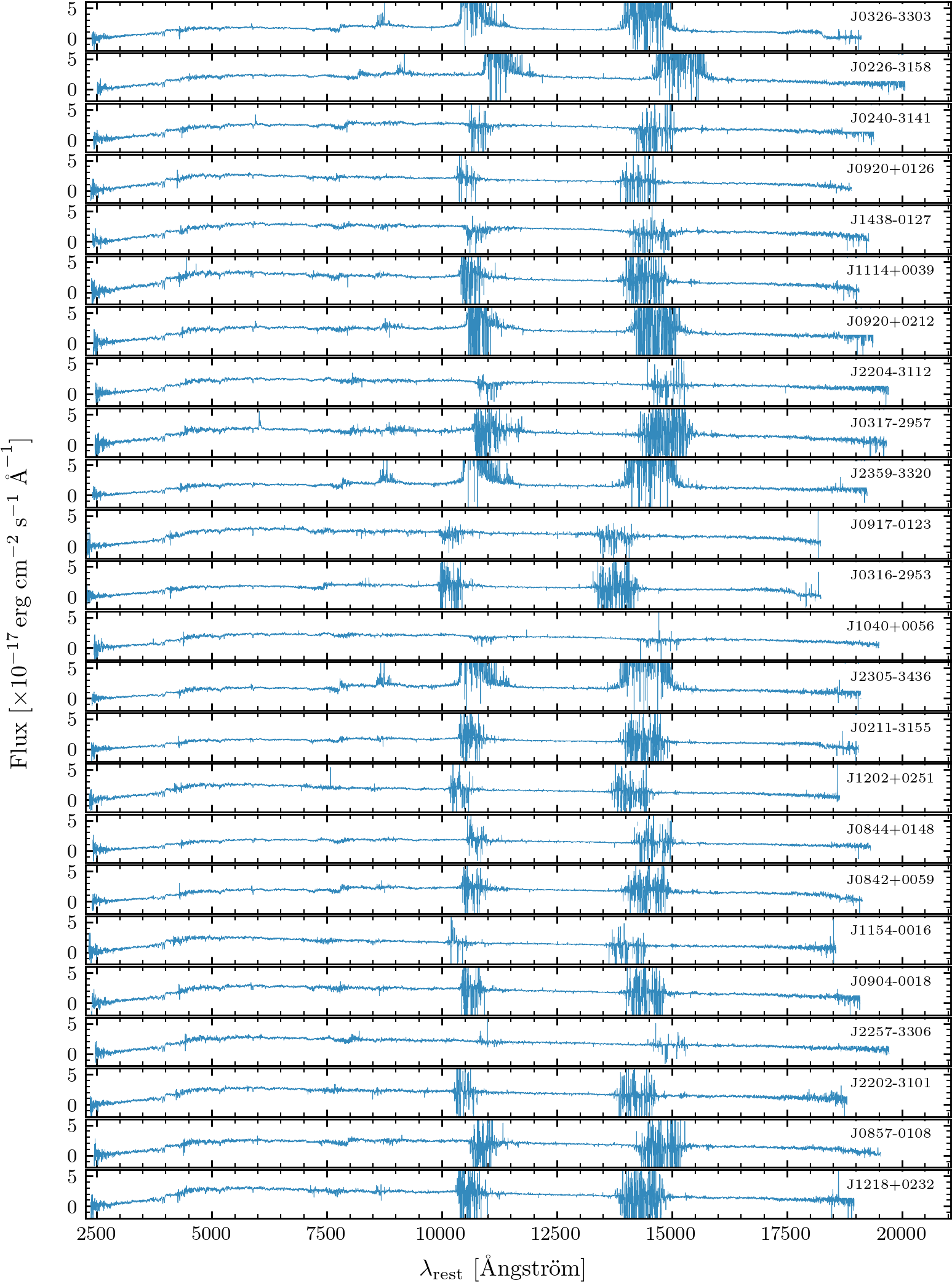}
    \caption{Same as the previous figure, but for the spectra in the medium SNR group ($25<\mathrm{SNR}<35$). }
\label{fig:spectra_medium}
\end{figure*}

In this Appendix, we show the 1D R50 spectra for all the galaxies in the \INSPIRE\ sample, ordered and divided by SNR group, as described in Section~\ref{sec:data} and Table~\ref{tab:spec_cat}. In particular, Figure~\ref{fig:spectra_high} shows the spectra with mean SNR $\mathrm{SNR}>35$, Figure~\ref{fig:spectra_medium} those with medium SNR ($25<\mathrm{SNR}<35$) and Figure~\ref{fig:spectra_high} those with SNR 
ranging from 15 to 25. For display purposes, the spectra have been smoothed with a median filter with a nine-pixel width. 
There are two slightly noisy regions (around 4200\AA, and around 7200\AA) where the three arms have been joined together, and two very noisy ones, which correspond to the large telluric bands. All these have been masked out from the analysis of the stellar kinematics and populations.  
We remind the reader that, for this DR, the stellar population parameters have been constrained from the spectral region [3500-7500]\AA. In this region, the SSP behaviour is solid and well studied, the influence of possible changes in the IMF slope and of very small ($<1\%$) star formation residuals are negligible. In future \INSPIRE\ papers we will focus more on the UVB and NIR regions.

\begin{figure*}
    \centering  \includegraphics[width=0.9\textwidth]{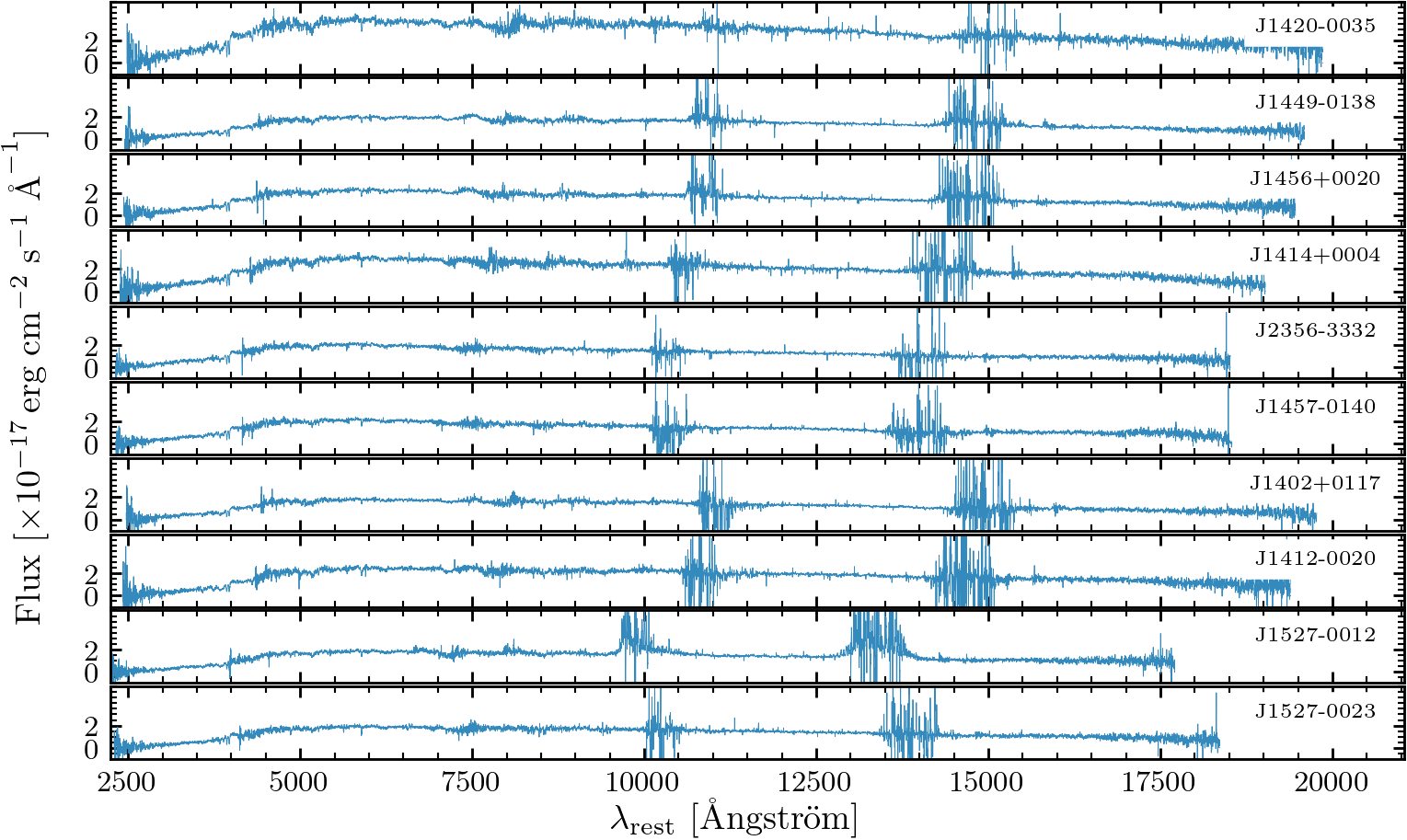}
    \caption{Same as the previous figure, but for the spectra in the low SNR group ($\mathrm{SNR}<25$).}
\label{fig:spectra_low}
\end{figure*}

%%%%%%%%%%%%%%%%%%%%%%%%%%%%%%%%%%%%%%%%%%%%%%%%%%

% Don't change these lines
\bsp	% typesetting comment
\label{lastpage}
\end{document}